\documentclass[twocolumn, times, trackchanges]{aastex63_bren}

\usepackage{CJK}
\usepackage{booktabs}
\usepackage{xcolor, soul}
\usepackage{amsmath, amssymb, bm}
\usepackage{graphicx}
\usepackage{dsfont}

\usepackage{float}

\newcommand{\teff}{T_{\rm eff}}
\newcommand{\logg}{\log g}
\newcommand{\xh}{[X/{\rm H}]}

\newcommand{\feh}{[{\rm Fe}/{\rm H}]}
\newcommand{\mgh}{[{\rm Mg}/{\rm H}]}
\newcommand{\mgfe}{[{\rm Mg}/{\rm Fe}]}
\newcommand{\afe}{[\alpha/{\rm Fe}]}
\newcommand{\xfe}{[{X}/{\rm Fe}]}
\newcommand{\xmg}{[{X}/{\rm Mg}]}

\newcommand{\expect}[1]{\left\langle {#1} \right\rangle}
\newcommand\kpc{{\rm kpc}}

\newcommand{\Xji}{X_{j,i}}

\newcommand{\sigjtot}{\sigma_{j,{\rm tot}}}
\newcommand{\sigjint}{\sigma_{j,{\rm int}}}
\newcommand{\sigjobs}{\sigma_{j,{\rm obs}}}
\newcommand{\sigktot}{\sigma_{k,{\rm tot}}}
\newcommand{\sigkint}{\sigma_{k,{\rm int}}}
\newcommand{\sigkobs}{\sigma_{k,{\rm obs}}}
\newcommand{\Cjkint}{C_{jk,{\rm int}}}

\newcommand{\Cjktot}{C_{jk,{\rm tot}}}

\newcommand{\Cjkobs}{C_{jk,{\rm obs}}}
\newcommand{\rhojkint}{\rho_{jk,{\rm int}}}
\newcommand{\rhojktot}{\rho_{jk,{\rm tot}}}
\submitjournal{ApJ}

\shorttitle{How Many Elements Matter?}
\shortauthors{Ting \& Weinberg}

\begin{document}
\begin{CJK*}{UTF8}{gbsn}

\title{How Many Elements Matter?}

\correspondingauthor{Yuan-Sen Ting}
\email{ting@ias.edu}

\author{Yuan-Sen Ting (丁源森)}
\affiliation{Research School of Astronomy \& Astrophysics, Australian National University, Cotter Rd., Weston, ACT 2611, Australia}
\affiliation{Research School of Computer Science, Australian National University, Acton ACT 2601, Australia}
\affiliation{Institute for Advanced Study, Princeton, NJ 08540, USA}
\affiliation{Department of Astrophysical Sciences, Princeton University, Princeton, NJ 08540, USA}
\affiliation{Observatories of the Carnegie Institution of Washington, 813 Santa Barbara Street, Pasadena, CA 91101, USA}

\author{David H. Weinberg}
\affiliation{Department of Physics, Department of Astronomy, and Center for Cosmology and Astro-Particle Physics,\\ The Ohio State University, Columbus, OH 43210, USA}
\affiliation{Institute for Advanced Study, Princeton, NJ 08540, USA}

%
%
%
%
%
%

\begin{abstract}
Some studies of stars' multi-element abundance distributions suggest at least 5-7 significant dimensions, but others show that many elemental abundances can be predicted to high accuracy from [Fe/H] and [Mg/Fe] (or [Fe/H] and age) alone.  We show that both propositions can be, and are, simultaneously true. We adopt a machine learning technique known as normalizing flow to reconstruct the probability distribution of Milky Way disk stars in the space of 15 elemental abundances measured by APOGEE. Conditioning on $\teff$ and $\logg$ minimizes the differential systematics. After further conditioning on [Fe/H] and [Mg/Fe], the residual scatter for most abundances is $\sigma_{[X/{\rm H}]} \lesssim 0.02$ dex, consistent with APOGEE's reported statistical uncertainties of $\sim 0.01 - 0.015$ dex and intrinsic scatter of $0.01-0.02$ dex. Despite the small scatter, residual abundances display clear correlations between elements, which we show are too large to be explained by measurement uncertainties or by the finite sampling noise. We must condition on at least seven elements to reduce correlations to a level consistent with observational uncertainties. Our results demonstrate that cross-element correlations are a much more sensitive probe of hidden structure than dispersion, and they can be measured precisely in a large sample even if star-by-star measurement noise is comparable to the intrinsic scatter.  We conclude that many elements have an independent story to tell, even for the ``mundane'' disk stars and elements produced by core-collapse and Type Ia supernovae.  The only way to learn these lessons is to measure the abundances directly, and not merely infer them.
\vspace{1cm}
\end{abstract}

%
%
%
%
%
%

\section{Introduction}
\label{sec:intro}

Ambitious Galactic spectroscopic surveys such as Gaia-ESO \citep{Gilmore2012}, APOGEE \citep{Majewski2017}, and GALAH \citep{Buder2020} have obtained high-resolution, high signal-to-noise ratio (SNR) spectra of hundreds of thousands of stars, spanning large swaths of the Milky Way disk, bulge, and halo and some nearby satellites such as the Sgr dwarf and the Magellanic Clouds. Other surveys including SEGUE \citep{Yanny2009}, RAVE \citep{Steinmetz2006}, and LAMOST \citep{Luo2015} have obtained lower resolution spectra of even larger stellar samples. The high-resolution surveys provide detailed chemical fingerprints for each program star, typically measuring 15-30 elements per star. This is further complemented by the lower resolution surveys which measure bulk metallicity and other abundance ratios \citep{Ting2017b, Xiang2019, Wheeler2020}. In concert with precise distances and proper motions from {\it Gaia} \citep{Gaia2018}, and with asteroseismic calibration of stellar ages \citep{Pinsonneault2018,Miglio2020}, these surveys afford an increasingly detailed picture of the Milky Way's stellar populations and dynamics, far beyond that available as recently as a decade ago.

It has long been recognized that the ratio of $\alpha$-elements (produced mainly by core-collapse supernovae) to iron peak elements (which are additionally produced by SNIa on a longer timescale) is an important dimension of stellar abundance variation in addition to overall metallicity characterized by $\feh$ (e.g., \citealt{Wallerstein1962,Tinsley1979,Fuhrmann1998,Bensby2003,Hayden2015}). However, the evidence on variations beyond $\feh$ and $\afe$ is mixed. On the one hand, \cite{Ness2019} found that a combination of $\feh$ and stellar age is sufficient to predict the value of other APOGEE $\xfe$ abundance ratios with precision comparable to the measurement uncertainties. In a related vein, \citet{Weinberg2019} and \citet{Griffith2021} found that an empirical ``two-process'' model fit to median abundance trends can predict the APOGEE $\xmg$ ratios for most disk and bulge stars to surprisingly high precision from $\mgh$ and ${\rm [Mg/Fe]}$ alone. On the other hand, by applying principal component analysis (PCA) to a much smaller, pre-APOGEE data set, \cite{Ting2012} found that five to seven components were needed to describe the multi-element abundances of solar neighborhood stars. \cite{Andrews2017} reached compatible conclusions with a different abundance sample. Working directly with spectra, \cite{Price-Jones2018} found that 10 components are needed to explain the diversity of APOGEE $H$-band spectra.

One goal of this paper is to reconcile these seemingly disparate conclusions and demonstrate that most elements contain critical information that cannot be simply inferred from the metallicity and $\alpha$-enhancement alone. Our approach is based on a powerful machine-learning technique called normalizing flow, which we use to create an accurate model of the probability distribution function (PDF) of 15 abundances measured in APOGEE disk stars, along with the stellar parameters $\teff$ and $\logg$. In particular, we can {\it condition} on the values of $\teff$, $\logg$, and a subset of elements, then evaluate the joint distribution of the remaining elements. The residual abundances --- star-by-star deviations from the conditional mean --- display significant cross-element correlations, revealing underlying structures.  Residual correlations only approach what we would expect from the observational uncertainties after conditioning on seven elements, signaling that most elements carry independent information.

Drawing on these results, we address the related question: is it still worth measuring many elemental abundances for large samples, even if $\feh$ and $\afe$ can already predict these abundances with $\sim 0.02$-dex root mean square (rms) dispersion? Our answer is an emphatic yes. Correlations can be measured at high significance in large samples even when the observational uncertainties are comparable to, or larger than, the star-to-star intrinsic dispersion of individual elements. This residual correlation structure from multi-element abundance trends may then provide critical diagnostics about myriad astrophysical processes, including stellar yields, ISM mixing, and merger history.

Conditioning the abundance distribution on Mg and Fe has some similarities to fitting a two-process model like that of \citet{Weinberg2019}. Weinberg et al.\ (\citeyear{Weinberg2021}; completed after the original version of this paper)  generalized this model-fitting approach and examine star-by-star correlated deviations from two-process predictions. While there is some overlap between these two papers, the two approaches have different underlying principles and complementary practical advantages. The normalizing flow method used here opens a novel route to minimizing observational systematics by conditioning on $\teff$ and $\logg$. It is conceptually and practically straightforward to condition on additional elements and thereby assess the independent information encoded in their abundances. \cite{Weinberg2021} aims to provide more physical insights about the emergence of these correlated deviations and their degree of correlation with Galactic location, stellar kinematics, and stellar age.

In this paper's next section we discuss why statistical correlations can reveal ``hidden'' degrees of freedom that might be buried in the dispersion about the conditional mean predictions. Section~\ref{sec:nflow} introduces the normalizing flow technique for describing arbitrary high-dimensional distributions. In \S\ref{sec:apogee} we apply this technique to a sample of disk red giants from APOGEE Data Release 16 (DR16; \citealt{Ahumada2020,Jonsson2020}). In \S\ref{sec:discussion} we discuss the implications of our methodology and results for the dimensionality of the stellar abundance distribution, for methods of abundance determination, for chemical tagging of co-natal stellar populations, and for the design of stellar spectroscopic surveys. In \S\ref{sec:conclusions} we summarize our findings and identify avenues for further application of these techniques.

%
%
%
%
%
%

\section{Variance, correlation, and dimensionality}
\label{sec:measuring}

As noted above, previous studies have shown that by conditioning on two elements representing core-collapse supernovae and SNIa, such as Mg and Fe, one can predict other elemental abundances to impressive accuracy. This does not necessarily imply that other elemental abundances are redundant. For example, if there are other residual correlations among groups of elements, it would mean that the elemental abundance space contains information beyond the amplitude of these two processes. These correlations, even if small, would imply that there are other hidden degrees of freedom (or ``dimensions'') in the Milky Way's chemical evolution. In this section we lay out the key ideas related to the measurement of such correlations before turning to the specifics of our method in \S\ref{sec:nflow} and \S\ref{sec:apogee}.

Consider $N$ random variables, $X_1, \cdots, X_N$, that represent $N$ elemental abundances, and let the correlation of any of the two variables be $\rho_{jk}$. By definition, the covariances of these $N$ variables are
\begin{equation}
    C_{jk} = \rho_{jk} \sigma_j \sigma_k~,
\end{equation}
where $\sigma_k$ is the standard deviation of the $k$-th variable. Suppose these $N$ variables represent elemental abundances after subtracting the mean abundances conditioned on Fe and Mg. For simplicity, we assume that these $N$ variables approach a multivariate Gaussian distribution. We would like to know if there are correlations among these residual abundances, indicating underlying physical structures in the abundance distribution.

\begin{figure*}
    \centering
    \includegraphics[width=1.0\textwidth]{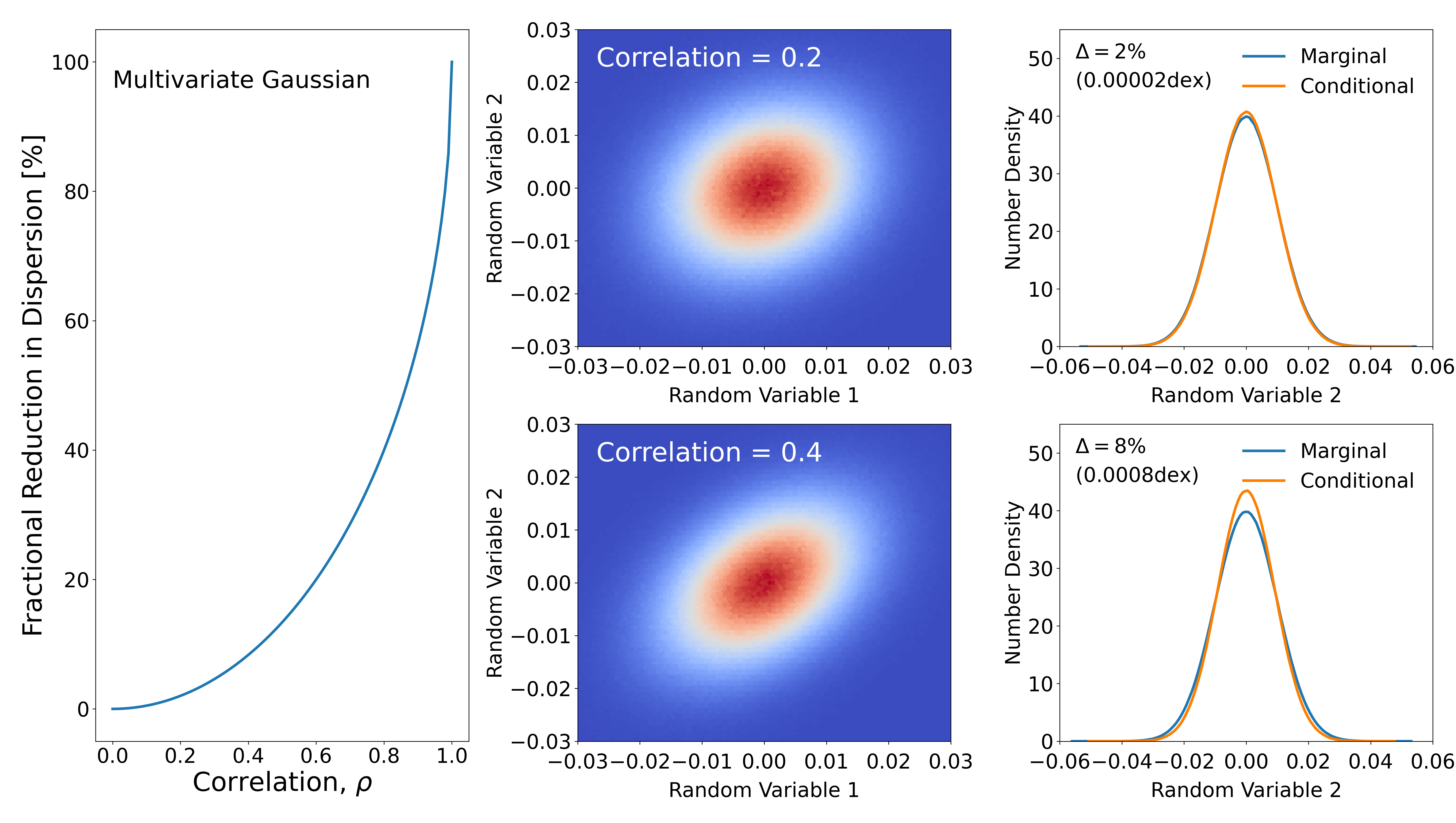}
    \caption{Measuring abundance correlation structure is more powerful than measuring the dispersion about the conditonal mean. {\it Left:} Fractional reduction in the dispersion of a random variable after conditioning on the value of a correlated variable, as a function of the correlation coefficient (Eq.~\ref{eq:scatter-reduction}). {\it Middle:} Density plot of random draws from a bivariate Gaussian distribution with dispersion of 0.01 dex for each variable (a typical intrinsic dispersion for APOGEE elemental abundances) and correlation coefficients $\rho = 0.2$ (top) and $0.4$ (bottom). {\it Right:} Distribution of variable 2 before (blue) and after (orange) conditioning on the value of variable 1. The changes of dispersion for these two values of $\rho$ are 2\% and 8\%, respectively, or less than 0.0008 dex. Correlations that are readily visible in the 2-d distributions of the middle panels are barely discernible in the change of dispersion.}
    \label{fig1}
\end{figure*}

One way to search for residual correlation is by investigating the change in variance\footnote{In this study, we will use the word ``dispersion'' to refer to the standard deviation of a variable, equal to the square-root of its variance.}, after conditioning on an additional variable. However, this method proves to be rather insensitive in practice. To understand this, we start with the simpler illustration shown in Fig.~\ref{fig1}. In the middle panels, we show two bivariate Gaussians with moderate strength correlations, $\rho =0.4$ and $0.2$. In the right panel we compare the variance of the second variable evaluated from the marginal distribution with the variance after conditioning on the first variable. Intuitively, if two variables are correlated, one would expect conditioning on the first to reduce the variance of the second, by taking advantage of the information provided by the first variable. However, Fig.~\ref{fig1} shows that a correlation that is easily detected visually in the middle panels is barely discernible when looking at the change of the variance as shown in the right panels.

This intuition can be formulated more rigorously. Assume $X_1$ to be the variable that we condition on. For a multivariate Gaussian, the new covariance matrix after conditioning on $X_1$ can be analytically calculated to be
\begin{equation}
    C'_{[2,N],[2,N]} = C_{[2,N],[2,N]} - C_{1,[2,N]} C_{11}^{-1} C_{[2,N],1}.
    \label{eq:conditioning}
\end{equation}
Each of these terms represents individual submatrices of the original covariance matrix $C_{jk}$, and we use the subscript $[2,N]$ to represent the ($N-1) \times (N-1)$ matrix for the second to the $N$-the variable. The diagonal entries of the new matrix $C'_{[2,N],[2,N]}$ are the variances upon the conditioning. By evaluating Eq.~\ref{eq:conditioning}, one can obtain an analytic expression for these new dispersion terms. The variance of the $k$-th variable can be written as
\begin{equation}
    \sigma^{\prime 2}_{k} = \sigma_k^2 (1 - \rho_{1k}^2).
\end{equation}
This implies that the fractional change of the dispersion is
\begin{equation}
    1 - \frac{\sigma_k'}{\sigma_k} = 1 - \sqrt{1-\rho_{1k}^2}~.
    \label{eq:scatter-reduction}
\end{equation}
For $\rho_{1k}^2 \ll 1$, the fractional change becomes $\simeq \rho_{1k}^2 / 2$.

The upper left panel of Fig.~\ref{fig1} illustrates this relation. For the specific correlation values of 0.2 and 0.4, the fractional changes of the dispersion are 2\% and 8\%, respectively. For an abundance intrinsic dispersion of 0.01 dex, typical of what we find for well measured APOGEE elements (see \S\ref{sec:apogee}), one would be looking for changes in the dispersion of $\sim 0.0008$ dex or less.

%
%
%
%
%
%

\subsection{Revealing abundance dimensionality through dispersion is challenging}
\label{sec:measuring_variance}

Although a reduction of 0.0008 dex might seem impossible to measure, we emphasize that, in principle, the residual variance of elemental abundances can be measured precisely in a large sample even if the observational uncertainties for individual stars are comparable to the intrinsic dispersion of the abundances. {\em The ``variance of the variance'' is not the variance itself.} Mathematically, suppose that we estimate the total variance for a given elemental abundance as
\begin{equation}
s_j^2 = {1 \over N_{\rm sample}} \sum_{i=1}^{N_{\rm sample}} \left(\Xji-\bar{{X}_j}\right)^2~,
\label{eqn:estimator}
\end{equation}
where $\bar{{X}}_j$ is the conditional mean abundance\footnote{We are taking $N_{\rm sample}$ large enough to ignore the slight bias caused by using the sample mean instead of the true mean.}. The mean-squared uncertainty of this estimate --- the variance of the variance --- can be calculated as follows.

For a Gaussian random variable, $(n-1)s_j^2/\sigjtot^2$ follows a chi-squared distribution with $(n-1)$ degrees of freedom, hence
\begin{eqnarray}
    {\rm Var} \, \frac{(n-1) s_j^2}{\sigjtot^2} = {\rm Var} \, \chi^2_{n-1} \\
    \Longrightarrow \quad \frac{(n-1)^2}{\sigjtot^4} {\rm Var} \, s_j^2 = 2(n-1).
    \label{eqn:v_of_v}
\end{eqnarray}
Here we have used the fact that the variance of $\chi^2_{n-1}= 2(n-1)$. We use the symbol $\sigjtot^2$ to represent the true total variance, and $s_j^2$ to represent the estimate of the total variance. For the $n= N_{\rm sample} \gg 1$, the rms fractional uncertainty in the variance is
\begin{equation}
    {\expect{(s_j^2-\sigjtot^2)^2}^{1/2} \over \sigjtot^2} = \left(2 \over N_{\rm sample} \right)^{1/2}~.
    \label{eqn:var_staterr}
\end{equation}
The equation shows that the variance of the variance can be small even if the variance itself is large.

In this study, we will mostly focus on the total measured variance without distinguishing its sources, but we note that the total variance of an elemental abundance is the sum of the intrinsic variance and the observational uncertainty,
\begin{equation}
\sigjtot^2 = \sigjint^2 + \sigjobs^2~.
\label{eqn:sigtot}
\end{equation}
Since the variance can be measured with a small fractional uncertainty in a large sample, as a corollary, the value of the intrinsic variance can be determined even if the observational uncertainty dominates the star-to-star dispersion.\footnote{Deriving the intrinsic dispersion does require accurate knowledge of the observational dispersion, and systematic uncertainties in the magnitude of this observational dispersion may dominate over statistical uncertainties. For example, if $\sigjint = \sigjobs$, then a systematic uncertainty of $\sqrt{2}$ in the value of $\sigjobs$ would be enough to explain all of the observed variance as a consequence of measurement dispersion.}

While we can in principle detect the {\it reduction} of variance from conditioning on a correlated variable with a large enough sample, it remains challenging in practice. For the examples in Fig.~\ref{fig1}, with $\rho=0.2$ and 0.4, the reductions in variance are 4\% and 16\% (twice the fractional reductions in the dispersion), and detecting them with $2\sigma$ statistical significance requires $N_{\rm sample} \simeq 5000$ and $N_{\rm sample} \simeq 300$, respectively. Although these numbers might appear within reach of large surveys, one must account for the fact that the effective number of stars after conditioning at a point in elemental abundance space is much smaller than the total size of the sample, a point we return to in \S\ref{sec:apogee} below. Because of that, as we will see, such a signal is often not statistically significant with the current data.

More generally, for $\rho^2 \ll 1$ the fractional reduction of the variance is $\rho^2$, and, from Eq.~\ref{eqn:var_staterr}, detecting this reduction at a significance of $\nu\sigma$ requires
\begin{equation}
    N_{\rm sample} \geq {2\nu^2 \over \rho^4}~.
    \label{eqn:Nmin_variance}
\end{equation}

%
%
%
%
%
%

\subsection{Detecting non-zero correlations directly is easier than detecting reductions in dispersions}
\label{sec:measuring_covariance}

While measuring the reduction of variance can be challenging, measuring non-zero correlations is much easier. For a multivariate Gaussian distribution, the correlation estimate's uncertainty due to finite sampling has a neat analytic approximation, known as the Fisher transformation. In particular, let the correlation be $\rho$ and its estimate be $r$, and let $z_{\rm fisher}= \tanh^{-1}(r)$. It can be shown that \citep{Fisher1921}, with a sample size $N_{\rm sample}$, the variable $z_{\rm fisher}$ is normally distributed with mean equal to $\frac{1}{2} \ln [(1+\rho)/(1-\rho)]$ and standard deviation of $1/\sqrt{N_{\rm sample} -3}$. In the null hypothesis with correlation $\rho=0 \simeq r$, we have $z_{\rm fisher} \simeq r$, and the mean of the distribution approaches 0 and the standard deviation $\simeq 1/\sqrt{N_{\rm sample}}$.

This result echoes Eq.~(\ref{eqn:var_staterr}), where the fractional uncertainty of the variance is $\sqrt{3/N_{\rm sample}}$. Detecting a non-zero correlation $\rho$ at significance of $\nu\sigma$ requires
\begin{equation}
    N_{\rm sample} \geq {\nu^2 \over \rho^2}~.
    \label{eqn:Nmin_correlation}
\end{equation}
Comparing Eqs.~(\ref{eqn:Nmin_correlation}) and~(\ref{eqn:Nmin_variance}), in addition to a factor of two gain, the key difference is that the denominator is $\rho^2$ instead of $\rho^4$.  Therefore, for $\rho^2 \ll 1$, it is far easier to detect a correlation of elements directly than to detect it through reduction of variance. For example, to detect a signal from $\rho =0.4$ at $2 \sigma$ significance only requires an {\em effective sample} of 25 stars, and for $\rho = 0.2$, 100 stars. We will see an empirical demonstration of this point in our APOGEE analysis.

A second advantage of focusing on cross-element correlations is that observational uncertainty is unlikely to produce artificial correlations at a significant level. We discuss this point further for the specific case of APOGEE abundance measurements in \S\ref{sec:apogee_aspcap} below, and we add a significant caveat in \S\ref{sec:apogee_aberration}. Nonetheless, the measurement and interpretation of correlations is not immune to observational uncertainty. The generalization of Eq.~(\ref{eqn:sigtot}) is

\begin{equation}
\Cjktot = \Cjkint + \Cjkobs~.
\label{eqn:cjktot}
\end{equation}
If the observational covariance is diagonal -- i.e., if the observational uncertainties are uncorrelated from one element to another -- then off-diagonal elements of the total covariance are just $\Cjktot = \Cjkint = \rhojkint \sigjint\sigkint$. For $j \neq k$, therefore, the relation between total and intrinsic correlations is
\begin{eqnarray}
    \rhojktot &=& {\Cjktot \over \sigjtot\sigktot} \\
    &=& \rhojkint {\sigjint\sigkint \over \sigjtot\sigktot} \\
    &=& \rhojkint \left(1+{\sigjobs^2\over\sigjint^2}\right)^{-1/2} \left(1+{\sigkobs^2\over\sigkint^2}\right)^{-1/2} ~.
    \label{eqn:rhojktot}
\end{eqnarray}
Thus, the measured correlations are always smaller in magnitude than the intrinsic correlations, by a factor that depends on the ratio of observational variance to intrinsic variance; for $\sigjobs \simeq \sigjint$ the reduction is about a factor of two. This reduction makes it more difficult to detect a given level of intrinsic correlation, especially correlations involving elements with large observational dispersion. In short, the measured correlation is a conservative limit, and if the measured correlation is significant then the detection is significant\footnote{Similar to the variance, systematic uncertainties in the magnitude of $\sigjobs$ may limit our ability to infer the true values of the intrinsic correlations. However, these systematic uncertainties cannot make a non-zero measured correlation consistent with a zero intrinsic correlation.}.

The third advantage of measuring cross-element correlations is that their information content is richer than that of excess variance alone, with informative clues to the physical origin of the residual abundance variations. Even if the magnitude of intrinsic correlation coefficients is uncertain, non-zero values of these coefficients are unambiguous evidence of remaining structure in the element distribution. Thus, the focus of our observational investigation will be to ask how many conditioning elements must be considered in the PDF before the residual correlations of the remaining elements are consistent with the observational uncertainties. For purposes of this paper, we take this number of conditioning elements to be our operative definition of the ``dimensionality'' of the disk stellar population in the space of APOGEE elemental abundances.

%
%
%
%
%
%

\subsection{Observational uncertainties in abundance measurements}
\label{sec:measuring_uncertainties}

The abundances of elements in the atmosphere of a star are estimated by fitting a model to the observed spectrum in which the value of the abundance is a free parameter. The modeling is complex, as it must include corrections for instrumental effects and telluric lines, a model of the stellar atmosphere, and a calculation of a synthetic spectrum from that atmosphere, which in turn depends on an adopted list of wavelengths and oscillator strengths for lines of different elements. Analyses of small sets of high-resolution spectra may be done ``by hand,'' but large surveys typically rely on codes that do automated spectral fitting and parameter optimization. For the giant stars in APOGEE DR16 that we will adopt in this study (see \citealt{Jonsson2020}), the APOGEE Stellar Parameters and Chemical Abundances Pipeline ({\sc aspcap}; \citealt{GarciaPerez2016,Holtzman2015}) assumes {\sc marcs} model atmospheres \citep{Gustafsson2008,Jonsson2020} and first fits a seven-parameter model\footnote{The parameters are $\teff$, $\logg$, the microturbulence $v_{\rm micro}$, and the abundance ratios [M/H], [$\alpha$/M], [N/M], and [C/M], where M is an overall metallicity scaling all elements together and [$\alpha$/M] scales the $\alpha$-elements O, Ne, Mg, Si, S, Ar, Ca, and Ti.} to the continuum-normalized $H$-band spectrum, then infers the abundances of other elements by varying [M/H] and fitting the spectrum in wavelength windows in which lines of that element dominate, with all other parameters held fixed.

There are two main sources of uncertainty in such estimates. The first is the statistical uncertainty arising from photon noise. The second is the systematic uncertainty arising because the models and calibrations used to infer abundances are imperfect. These systematic effects can include departures from LTE or plane-parallel geometry, incomplete or inaccurate line lists, and data-related effects such as inaccurate line-spread functions or incomplete removal of telluric contamination. In an absolute sense --- the difference between a star's estimated and true abundances --- the systematic uncertainties are often larger than the statistical uncertainties in a high-SNR spectroscopic survey. For this reason, the high statistical precision of abundance uncertainties derived from high-SNR spectra is often dismissed as unrealistic. However, to the extent that the systematic uncertainties are the same for all sample stars, {\it }they do not add dispersion to the measured abundances. Observational contributions to element dispersion arise from photon noise, and they can also arise from {\it differential} systematic uncertainties within the sample.

The latter contribution can be minimized by examining stars within a narrow range of parameters such as $\teff$, $\logg$, and $\feh$, or, with the method implemented in this paper, by examining distributions conditioned on these parameters. One of our study's innovations is the ability to minimize the systematic uncertainties through flexible modelling of the elemental abundance space, conditioned on variables that contribute to the systematic uncertainties. This innovation allows the observational dispersion to approach the photon noise limit, enabling a statistically significant detection of the residual abundance correlations.

Accurate characterization of statistical abundance uncertainties due to photon noise is therefore important even if the absolute abundance uncertainty is dominated by systematic uncertainties. The impact of photon noise on abundance measurements is primarily diagonal, i.e., adding observational uncertainties that are statistically independent for different elements. However, photon noise uncertainties are not entirely diagonal because uncertainties in some parameters (especially $\teff$, $\logg$, and overall metallicity) affect the models used to infer all abundances, and additionally because some abundances may be measured from blended spectral features or from molecular lines that involve two elements. We estimate these off-diagonal uncertainties for APOGEE abundances in \S\ref{sec:apogee_aspcap} below.

APOGEE reports statistical uncertainties based on repeat observations of a subset of stars, which are used to derive an empirical formula relating the standard deviation of repeat observations to a star's $\teff$ and [M/H] and the SNR of its spectrum \citep{Jonsson2020}.  These uncertainties are usually larger than those derived from the $\chi^2$-fitting of the spectrum, which implies that variations of observing conditions are contributing to the observational uncertainties in addition to Poisson fluctuations in the number of photons per pixel.  Nonetheless, we will generally refer to these statistical measurement uncertainties with the shorthand phrase ``photon noise.''

\begin{figure*}
    \centering
    \includegraphics[width=1.0\textwidth]{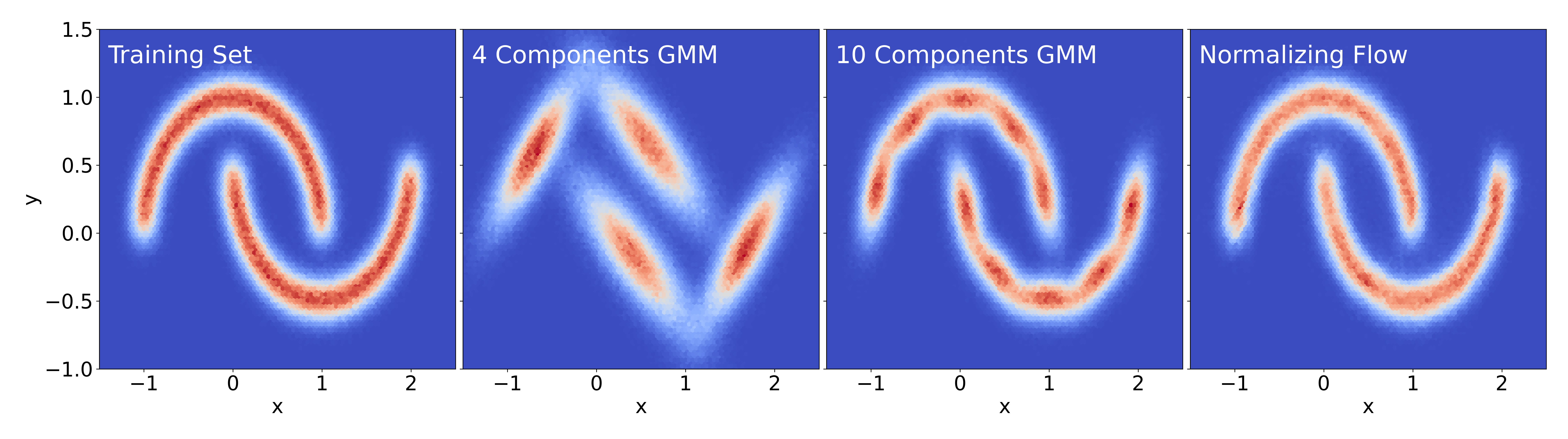}
    \caption{Normalizing flow can faithfully describe challenging distributions. As an illustration, we generate a 2D moon-shaped PDF and plot its density distribution (via a sample of $10^6$ random draws), as shown in the left panel. The two middle panels show the recovery of the distribution assuming 4-component and 10-component Gaussian Mixture Models. These fail to describe the distribution, even in 2D. On the right, we demonstrate a much better recovery with a normalizing flow, as illustrated further in Fig.~\ref{fig3}.}
    \label{fig2}
\end{figure*}

%
%
%
%
%
%

\section{Describing Distributions with Normalizing Flows}
\label{sec:nflow}

Elemental abundance space correlations can reveal many subtle properties about stellar yield processes and ISM mixing. In particular, if we can recover the multi-dimensional distribution $p(\xh)$ spanned by the elemental abundances of stars, we can then calculate moments of the distribution and correlations between elemental abundances. In practice, the data that we collect (e.g., from APOGEE) is only an ensemble of realizations drawn from the PDF, $\{\xh_i\}$, where each realization $i$ consists of the measured abundances of an individual star. Therefore, the first step to study the elemental abundance space requires tools that can faithfully recover the PDF $p(\xh)$ from the ensemble of realizations $\{\xh_i\}$, which we will elaborate on in this section.

One way of recovering a distribution from an ensemble of realizations is to conjecture a functional form for the distribution and then maximize the likelihood of the parameters. Mathematically, let $\vec{x}$ be an $N$-dimensional random variable, and let $\{ \vec{x}_i \}$ be the ensemble of the realizations. If we assume $f_\theta(\vec{x})$ to be the functional form of the normalized PDF, characterized by $\theta$, finding that distribution that best describes the data translates into a simple question of finding the $\theta^*$ that optimizes the likelihood,
\begin{equation}
    \theta^* = {\rm arg max}_\theta \Big[ \sum_i \ln f_\theta (\vec{x}_i) \Big]~.
\end{equation}
However, for an arbitrary distribution, our human heuristic on the functional form $f_\theta$ can be quite limiting (see Fig.~\ref{fig2}). Especially for a high-dimensional and irregular distribution like that of stars in elemental abundance space, any ad hoc functional form might not fully capture all the distribution details. This is where the idea of normalizing flow, a machine learning tool that is rapidly growing in applications, can play an important role.

The basic idea of normalizing flow is to use neural networks, characterized by the neural network coefficients $\psi$, as a change of variables. More specifically, the goal is to transform the multi-dimensional random variable $\vec{x}$ to a new random variable $\vec{z} = f_\psi(\vec{x})$ of the same dimension, such that $p(\vec{z})$ is a much simpler and more recognizable distribution than $p(\vec{x})$. We call the distribution $p(\vec{z})$ the ``base distribution.'' In this study, the base distribution is chosen to be a unit-multivariate Gaussian distribution with a zero mean and an identity covariance matrix.

For a change of random variables, one needs to take into account the change of measure $f'_\psi(\vec{x})$, known as the Jacobian. More precisely,
\begin{equation}
  p(\vec{z}) \, {\rm d}{\vec{z}} = p(f_\psi(\vec{x})) |f'_\psi(\vec{x})| \, {\rm d}{\vec{x}} .
\end{equation}
Therefore, in order to evaluate likelihood in the $\vec{z}$ space, we need to ensure that the neural networks' Jacobian $|f'_\psi(\vec{x})|$ is easily and analytically calculatable. With that premise, we can then optimize for the neural network coefficients $\psi$, such that the neural network transforms the original ensemble $\{\vec{x}_i \}$ to $\{ \vec{z}_i \}$ and that the ensemble of $\{ \vec{z}_i \}$ approaches a multivariate Gaussian distribution. Mathematically, we optimize for the neural network coefficients $\psi$ through the standard back-propagation technique with a rectified ADAM optimizer \citep{Liu2019}, and find $\psi^*$ such that
\begin{equation}
  \psi^* = {\rm arg max}_\psi \Big[ \sum_i \ln P_{\vec{z}}(f_\psi(\vec{x}_i)) |f'_\psi(\vec{x}_i)| \Big]
\end{equation}
In other words, we maximize the likelihood such that the ensemble of $\vec{z}$ approaches a unit-Gaussian distribution. We emphasize that the condition that the Jacobian of the neural networks be analytically calculable is crucial, since it would be otherwise prohibitively expensive to perform the optimization.

Besides the Jacobian criterion, another criterion is equally important. Note that $f_\psi$ transforms the random variable from $\vec{x}$ to $\vec{z}$, with which we calculate the likelihood. Yet, the multivariate Gaussian distribution $p(\vec{z})$ is the one from which we can easily sample. Therefore, if we want to sample $p(\vec{x})$ effectively, the neural network $f_\psi$ has to be analytically invertible so that we can first sample $\vec{z}$ from $p(\vec{z})$, then evaluate $f_\psi^{-1}(\vec{z})$ to attain an ensemble of $\vec{x}$.

\begin{figure*}
    \centering
    \includegraphics[width=1.0\textwidth]{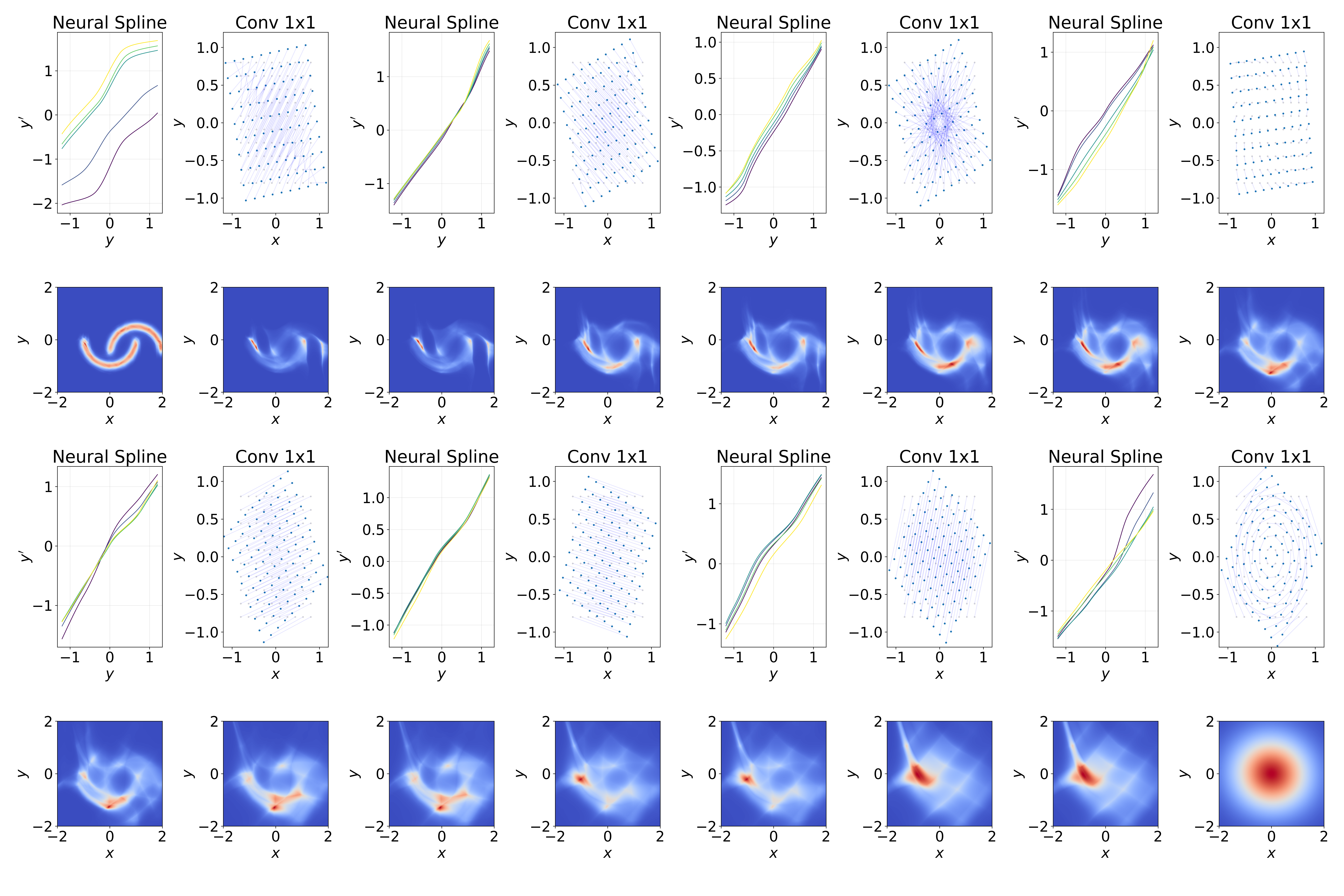}
    \caption{Structure of the normalizing flow used in Fig.~\ref{fig2} and this study. Normalizing flows adapt neural networks as a change variable. The goal is to transform the distribution from a complicated distribution (top left) to a simple base distribution (bottom right) through a series of gradual transformations. Here we assume a unit-multivariate Gaussian as the base distribution. This study adopts a normalizing flow that comprises eight units of Conv1x1 and Neural Spline Flow coupling. The Conv1x1, as illustrated in the even columns, performs linear transformations on the variables, while Neural Spline Flow, as shown in the odd columns, transforms the variables through spline interpolations parametrized by neural networks. The odd rows demonstrate the transformations, and the even rows show the results of these gradual transformations (see text for details).}
    \label{fig3}
\end{figure*}

In short, to use neural networks as a flexible change of random variables, the networks need to satisfy two criteria, namely (a) the Jacobian of the network is analytic and (b) the neural network is analytically invertible. The subset of neural networks that satisfy these two criteria are known as ``normalizing flows.'' They can transform any random variable from a complex distribution into a unit-Gaussian (``normal'') distribution (see Fig.~\ref{fig3}).

The idea of normalizing flow has inspired the machine learning community since its inception \citep[e.g.,][]{JimenezRezende2015}. The ability to describe high-dimensional PDFs given any ensemble also makes it one of the most flexible machine learning tools to apply to the physical sciences. However, normalizing flow has had a slow start in astronomy, with only a handful of applications to date. For example, \citet{Cranmer2019} applied normalizing flow to describe the color-magnitude diagram of the Gaia data, and \citet{Reiman2020} used normalizing flow to model quasar continua. More recently, \citet{Green2020} adopted normalizing flows to describe phase-space distributions and solve for Galactic dynamics.

In this study, we will adopt normalizing flows to describe the elemental abundance distribution of stars. We assume a similar normalizing flow to that adopted by \citet{Green2020}. More specifically, we adopt eight units of ``Neural Spline Flow'' coupled with the ``Conv1x1'' operation (or ``GLOW'' in the machine learning lingo). Each Neural Spline Flow unit consists of three layers of densely connected networks with 16 neurons. We refer interested readers to the original articles \citep{Kingma2018,Durkan2019}. Here we give a brief overview.

Roughly, a Neural Spline Flow performs an invertible spline transformation whose Jacobian is analytically calculable. The Conv1x1 operation, on the other hand, performs a linear transformation of the variables. Like most normalizing flows, the trick to ensure an analytic Jacobian is through the idea of ``coupling'', i.e., performing a change of variable for a subset of variables at each transformation unit. By only changing a subset of the variables each time, we can ensure that the $f'_\psi$ matrix is triangular, which allows for a more straightforward calculation of the Jacobian (the determinant of $f'_\psi$). When many transformation units (eight in total, in our study) are applied together, and each unit transforms a subset of variables in a tractable way, we can change the complex multivariate random variables gradually to become a simpler Gaussian distribution (see Fig 3).

We chose to use eight units of normalizing flows as this architecture is sufficient to capture the distribution of the elemental abundance space. Using more units could, in principle, describe the distribution better but with a higher computational training cost. From our bootstrapping experiments (see Section~\ref{sec:apogee_sampling}), we found that any systematics incurred by our architecture are subdominant compared to the finite sampling noise. We will come back to this point in Section~\ref{sec:survey-design}.

Since the idea of normalizing flow is rather new, well-tested public codes are limited. Through our own extensive exploration, we found that most public packages do not seem to perform as well as hand-crafted codes. We therefore use our own codes adapted from Github repository {\sc karpathy/pytorch-normalizing-flows}. A similar code was also used in \citet{Green2020} and is publicly available on Github\footnote{https://github.com/tingyuansen/deep-potential}.

To demonstrate the power of normalizing flow, in Fig.~\ref{fig2}, we present a case study with a simple double moon-shaped distribution. We chose this toy example as it loosely resembles the elemental abundance space distribution of the high-$\alpha$ population versus the low-$\alpha$ population. The left panel shows the ensemble of $10^6$ realizations drawn from the moon-shaped distribution. The two middle panels illustrate the cases where we attempt to describe the distribution with Gaussian Mixture Models, with four and ten components, respectively. Even in this simple example of 2D variables, Gaussian Mixture Models fail to represent the actual distribution faithfully. The right panel shows the results fitted with a normalizing flow, which clearly outperforms the Gaussian Mixture Models in depicting the double moon-shaped distribution.

We further demonstrate how normalizing flow works in Fig.~\ref{fig3}. From left to right, and from top to bottom, Fig.~\ref{fig3} shows how normalizing flow gradually transforms the complicated double moon-shaped distribution to the simple base distribution (a 2D Gaussian) through a series of operations. The alternate panels show the neural spline flow operation and the Conv1x1 operation. For each step, the top panel shows the transformation, and the bottom shows the result of the transformation. For the Conv1x1 transformation, in this 2D case, the operation essentially redefines the axes by rotating the axes. On the top panels, the black points show a regular grid in the original dimensions, and the blue points show the transformed locations after the rotation. We connect the blue and black points to visualize the rotation. As for the neural spline flow operation, it performs a 2D invertible spline. To visualize this transformation, the lines in different colors on the top panels are initiated from different original x values (from purple to yellow, the x value ranges from -1 to 1, with a spacing of 0.5), and the panels show the transformed y values as a function of the original y values after the spline operation.

At its core, the ability to invert a neural network stems from that fact that we can design a network $f_\psi({\vec{x}}) = f(s({\vec{x}}))$, where $s$ is a neural network, such that the inverse of $f_\psi({\vec{x}})^{-1} = g(s({\vec{x}}))$ can be written as a function of $s$ as well. In this way, we can explicity invert the $f_\psi({\vec{x}})$ without retraining any network.\footnote{To demonstrate how this can be done, we will explain with one of the simplest forms of normalizing flows -- RealNVP \citep{Dinh2017}. In RealNVP, the normalizing flow transformation can be written as $\vec{y}_{1:d}$ = $\vec{x}_{1:d}$ and $\vec{y}_{d+1:D} = \vec{x}_{d+1:D} \odot \exp(s_1(\vec{x}_{1:d})) + s_2(\vec{x}_{1:d})$, where the subscript denotes the subdimension, $s_1$ and $s_2$ are two trainable network, and $\odot$ is the element-wise product. It is easy to verify that the inverse of this operation, as well as the Jacobian, can be explicitly written as a function of $s_1(\vec{y})$ and $s_2(\vec{y})$. Therefore, once we find the best $s_1$ and $s_2$ through the ``forward" network, the ``inverse" network is automatically defined. The trick here depends on what is known as ``coupling," i.e., holding some of the dimensions invariant (in the notation above, the first $d$ dimensions). Although for individual ``units," some dimensions are not transformed, by having more than one unit in the normalizing flow, and holding different dimensions invariant for the different units, one can achieve transformation for all the dimensions while ensuring invertibility. For interested readers, we recommend \citet{Weng2018} for an introductory text on how to design other multidimensional neural networks that can be explicitly inverted.} With this in mind, depicting a conditional distribution $p(\vec{x}|\vec{y})$ with normalizing flow only requires a minor modification of what we have discussed thus far. Instead of the usual change of variable, characterized by a neural network $f_\psi$, it suffices to ensure that the neural network coefficients also depend (continuously) on the conditioning variable $\vec{y}$. In other words, instead of $f_\psi$, we have $f_{\psi(\vec{y})}$. Operatively, for any function $h(\vec{x})$ in a Neural Spline Flow, where $h:\mathbb{R}^k \rightarrow \mathbb{R}^k$ is a function that transforms some part of the random variables with $k$ dimension, to build in the dependency on the $m$-dimensional variable $\vec{y}$ we consider $h(\vec{x},\vec{y}) = h_1 (\vec{x}) h_2( \vec{y} )$, where $h_1: \mathbb{R}^k \rightarrow \mathbb{R}^k$ is the mapping that does the original transformation, and $h_2: \mathbb{R}^m \rightarrow \mathbb{R}^k$ is an additional function that governs how the mapping $h_1$ should be modified with different $\vec{y}$. Both $h_1$ and $h_2$ are each represented by a densely connected network with three hidden layers and 16 neurons each. To put it simply, for different conditioning variables $\vec{y}$, the neural network depicts a different change of variable $f_{\psi(\vec{y})}$ that maps the variables $\vec{x}$ (with the same $\vec{y}$) to variables $\vec{z}$ that form a unit-Gaussian distribution. In principle we could compute a conditional PDF by numerically integrating the joint PDF, but training a new normalizing flow by this modified technique is a more efficient way to attain an accurate conditional PDF and to evaluate the conditional likelihood.

In this study, we will adopt the conditional normalizing flow to model the distribution of the APOGEE elemental abundances. As discussed earlier, the conditioning variables $\vec{y}$ include $\teff, \logg$ of the stars, as well as a subset of elemental abundances (such as Fe and Mg). The independent variable $\vec{x}$ is comprised of other elemental abundances that we do not condition on. When we condition on a new set of variables, we retrain the normalizing flow each time\footnote{Training a normalizing flow with $\sim$20{,}000 data entries in this study takes about an hour on a single CPU core. In this study with 17 dimensions (15 elemental abundances, $\teff$, and $\logg$), any improvement from the GPU acceleration appears to be minimal.}. Recall that normalizing flow allows us to sample $p(\vec{x}|\vec{y})$ through the inverse mapping $f_{\psi(y)} : (\vec{z},\vec{y}) \rightarrow \vec{x}$. Therefore, for any specific reference value of the conditioning variable $\vec{y}$, we can evaluate the correlation of the independent elemental abundances $\vec{x}$ by drawing samples from $p(\vec{x}|\vec{y})$. As we will see in the next section, the ability to draw samples and evaluate the correlation matrix at any values for the conditioning variables will come in handy for several aspects of this study.

\begin{figure*}
    \centering
    \includegraphics[width=1.0\textwidth]{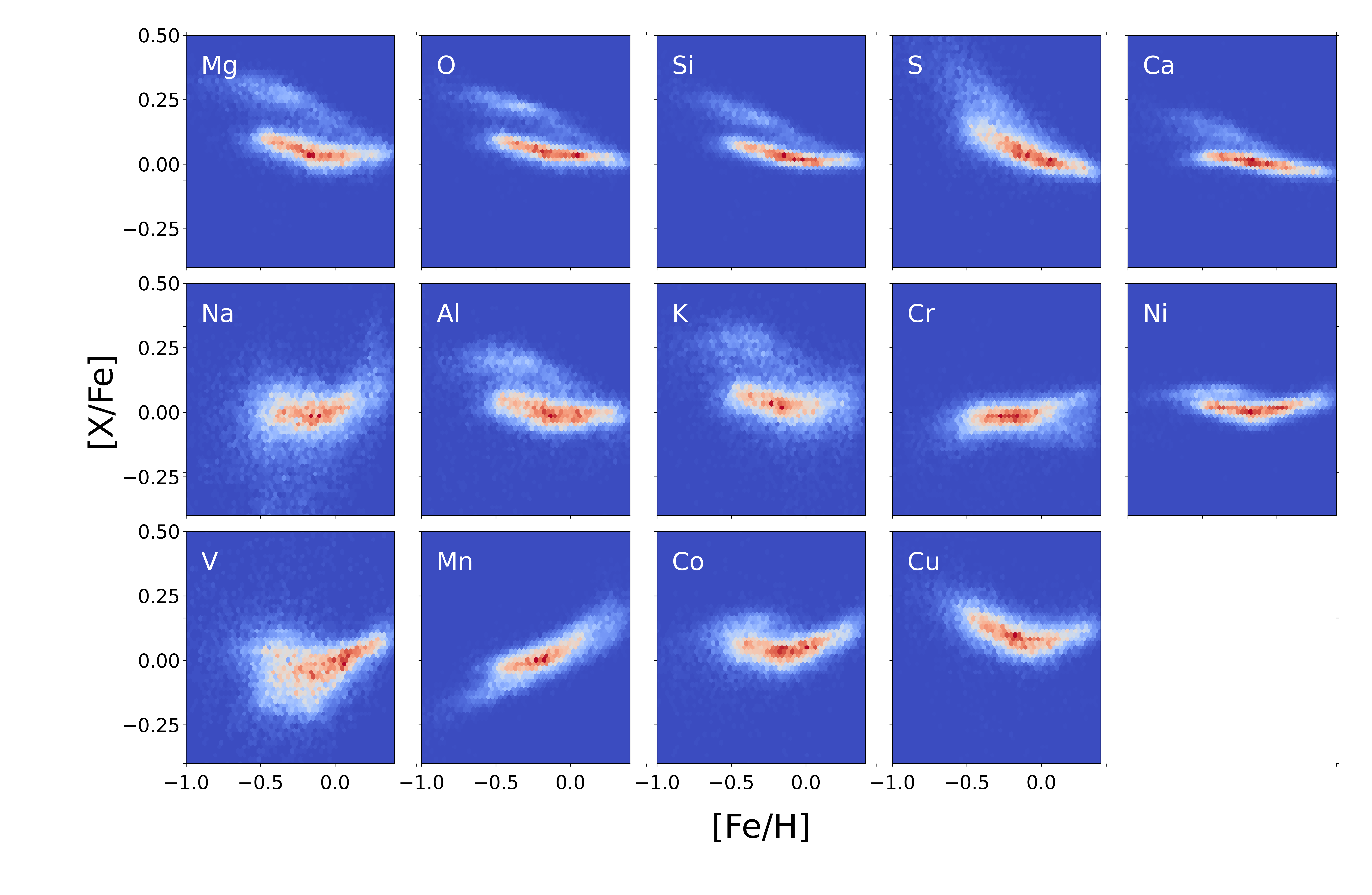}
    \caption{Abundance distribution in $[X/{\rm Fe}]$ versus ${\rm [Fe/H]}$ of the APOGEE training set. We adopt the training set to measure the conditional PDF and pairwise correlations of the 15 elemental abundances. The training set comprises $20{,}111$ APOGEE disk stars with stellar parameters $4100 {\rm K} < \teff < 4600 {\rm K}$ and $1 < \logg < 2.5$. We restrict the parameter range of our training set to minimize differential uncertainties in the abundances, and we also restrict the sample to SNR$\, > 200$ for $\mgh\geq -0.5$; SNR$\, > 100$ for $\mgh < -0.5$ to ensure that the abundance measurements are robust.}
    \label{fig4}
\end{figure*}

\begin{figure*}
    \centering
    \includegraphics[width=1.0\textwidth]{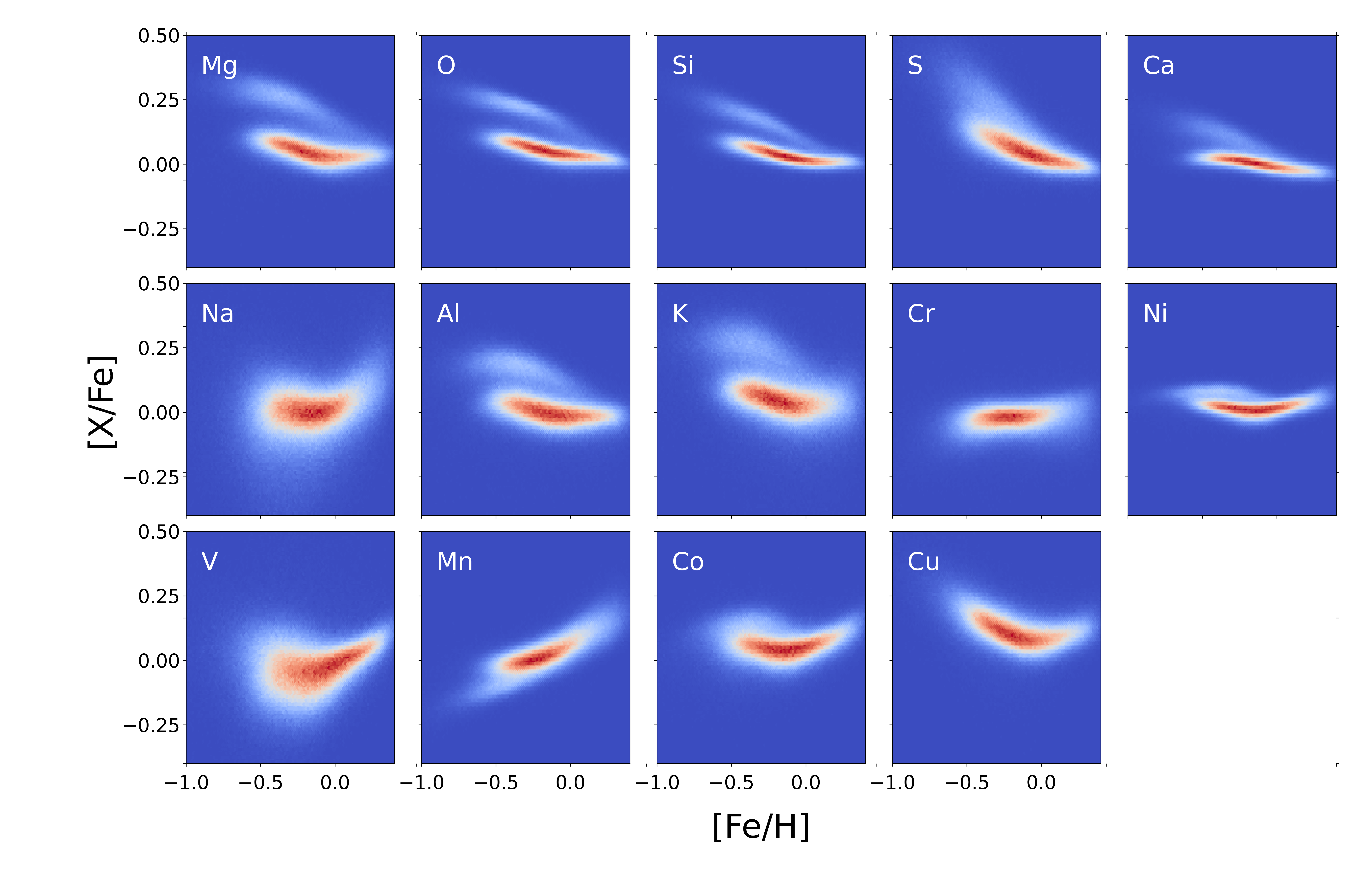}
    \caption{Distributions of $\xfe$ vs. $\feh$ from a sample of $10^6$ ``stars'' drawn from the joint PDF constructed by applying normalizing flow to the APOGEE training set shown in Fig.~\ref{fig4}. Comparison with Fig.\ref{fig4} demonstrates that the normalizing flow describes the APOGEE abundance PDF faithfully. The distributions in this figure are smoother because of the much larger number of drawn samples.}
    \label{fig5}
\end{figure*}

\begin{figure*}
    \centering
    \includegraphics[width=1.0\textwidth]{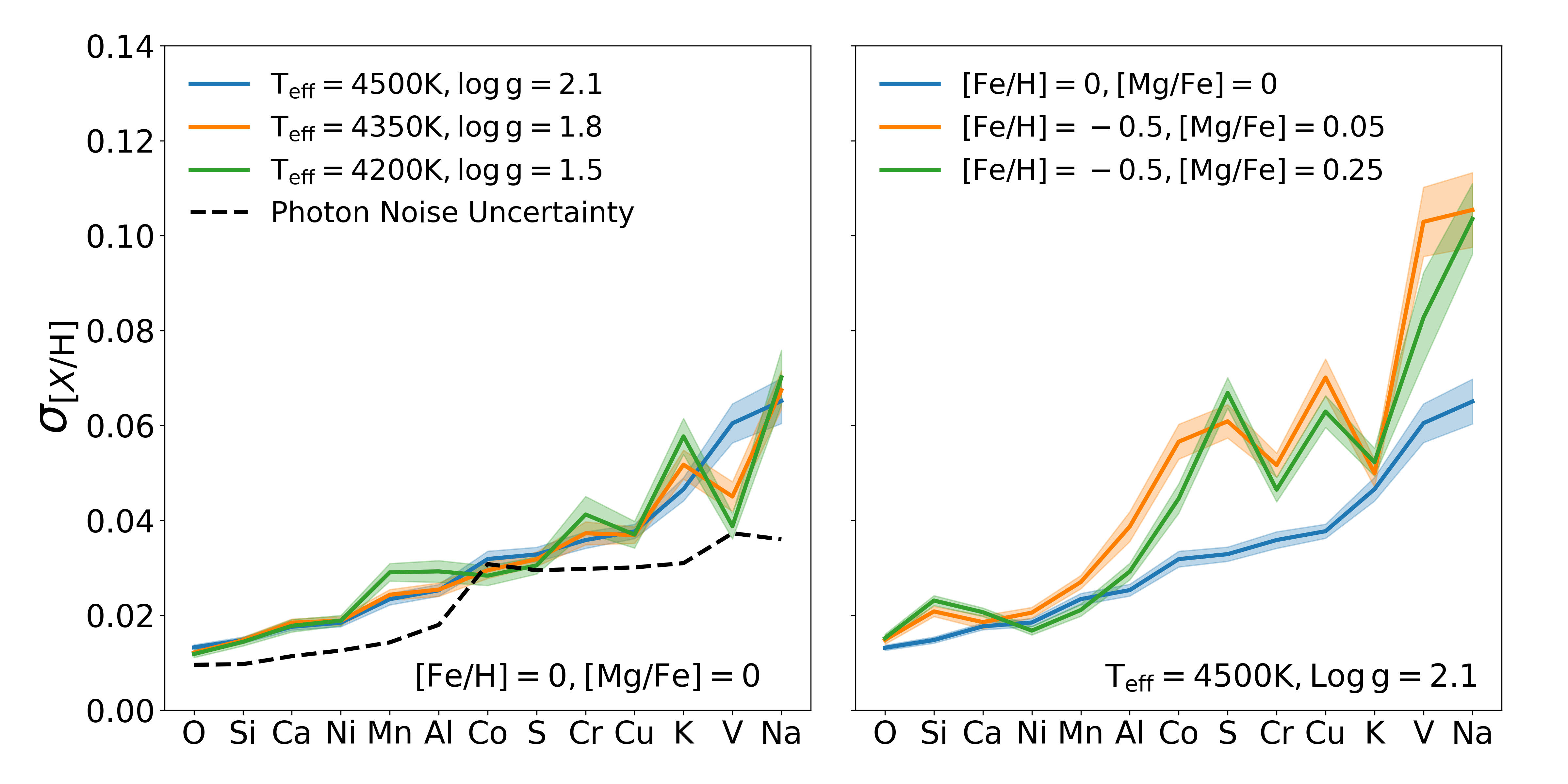}
    \caption{Dispersion of APOGEE elemental abundances about the conditional mean predicted at fixed stellar parameters, $\feh$, and $\mgfe$, computed from the conditional PDF of our normalizing flow model trained on the APOGEE disk star sample. We define the dispersion $\sigma_{\xh}$ to be half the difference between the 16th- and 84th-percentile abundance values in the marginal PDF for each element. In the left panel, blue, green, and orange lines show the dispersion at three different reference points of $\teff$ and $\logg$ as labeled, all for $\feh=\mgfe=0$, and bands indicate the finite sampling uncertainty inferred from bootstrap realizations. The dashed line shows the mean photon noise uncertainty reported by {\sc aspcap} for each element. In the right panel, green and orange lines show the dispersion at two other reference points of $\feh$ and $\mgfe$ as labeled, all for $\teff=4500\,{\rm K}$, $\logg=2.1$.}
    \label{fig6}
\end{figure*}

\begin{figure*}
    \centering
    \includegraphics[width=1.0\textwidth]{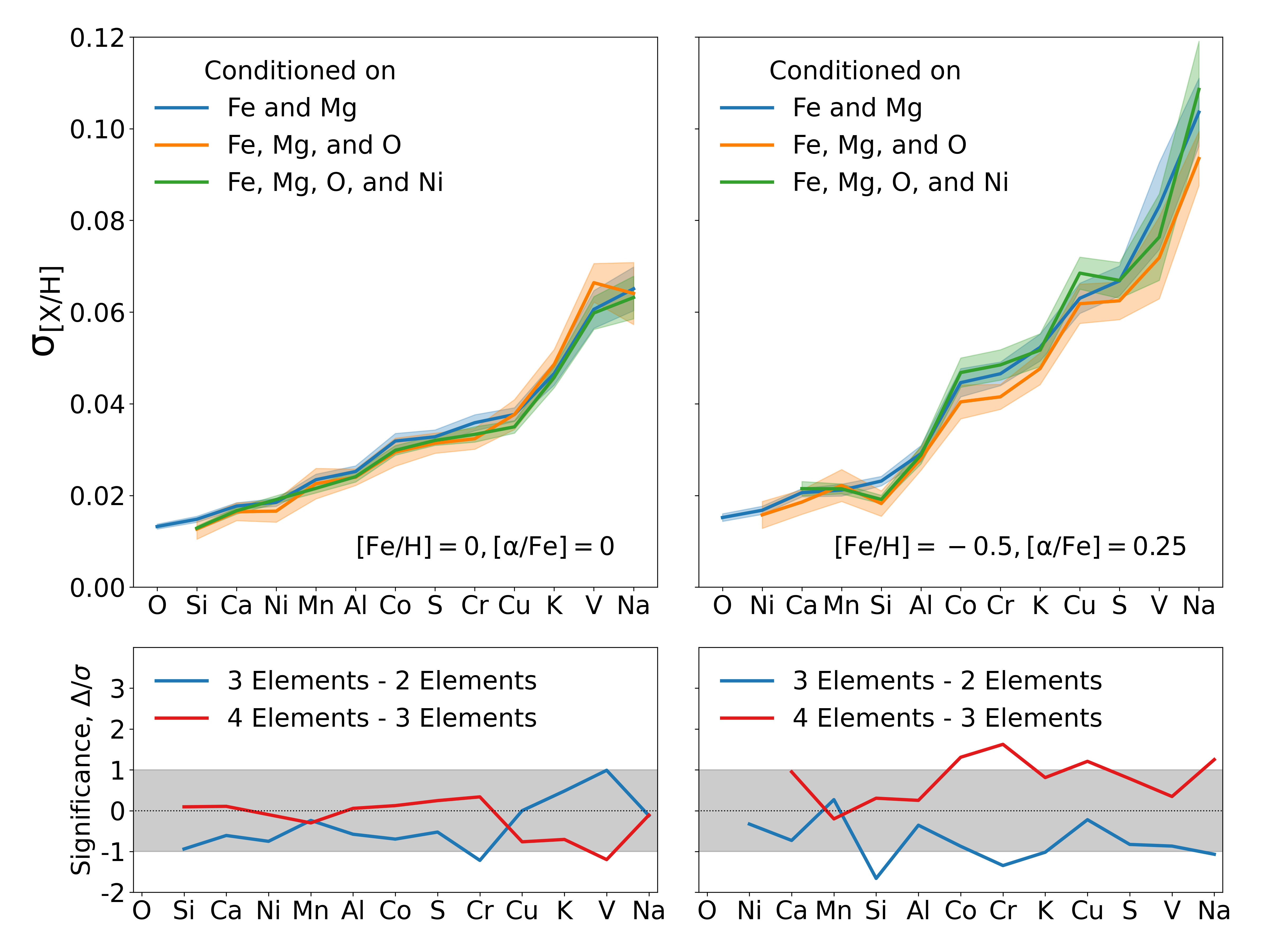}
    \caption{Dispersion of abundances about the conditional mean after conditioning on two, three, or four elements (in addition to $T_{\rm eff} = 4500\,$K and $\logg = 2.1$). Left and right panels show two different conditional locations in metallicity and $\alpha$-enhancement, as labeled. Lower panels show the change in dispersion in units of the finite sampling uncertainty inferred from the standard deviation of bootstrap samples of the data set. The reduction of dispersion from adding conditional elements beyond Fe and Mg is statistically insignificant in this data sample, but the reductions in residual correlations are much more detectable as shown below in Fig.~\ref{fig10} (see also theoretical arguments in \S\ref{sec:measuring} for better intuition).}
    \label{fig7}
\end{figure*}

%
%
%
%
%
%

\section{How many elements matter?}
\label{sec:apogee}

%
%
%
%
%
%

\subsection{Data sample}
\label{sec:apogee_data}

We select Milky Way disk stars from APOGEE DR16 with the following selection criteria:
\begin{itemize}
\item Galactocentric radius $3 \kpc \leq R \leq 13 \kpc$,
\item Midplane distance $|Z| \leq 2\kpc$,
\item $1 \leq \logg < 2.5$,
\item $4100\,{\rm K} \leq \teff < 4600\,{\rm K}$,
\item $-0.75 \leq \mgh < 0.45$
\item ${\rm SNR} > 200$ for $\mgh\geq -0.5$; ${\rm SNR} > 100$ for $\mgh < -0.5$.
\end{itemize}
We eliminate stars with {\sc star\_bad}, {\sc extratarg}, or {\sc no\_aspcap\_result} flags set, or with flagged values of $\feh$ or ${\rm [Mg/Fe]}$. Our geometric cuts are based on distances from {\sc AstroNN} \citep{Leung2019a,Leung2019b}, which are publicly available as a value-added catalog for APOGEE DR16.

The $\logg$ range selects luminous giants, allowing us to sample the full range of Galactocentric radii. The lower $\teff$ cut eliminates cool stars for which {\sc aspcap} abundances may be less reliable. Further, many stars with $<4100\,$K in the parent sample have flagged values of Mn and Cu. The upper $\teff$ cut eliminates red clump stars (see fig.~3 of \citealt{Vincenzo2021,Pinsonneault2018}), which should have reliable abundances but could be offset from giant branch stars (see fig.~13 of \citealt{Jonsson2020}). More generally, our use of restricted $\logg$ and $\teff$ ranges is intended to reduce the impact of differential systematic uncertainties (see \S\ref{sec:apogee_baseline}); we further mitigate the differential systematics by conditioning on $\teff$ and $\logg$ when training the normalizing flows. Our high SNR threshold is intended to select stars with the most reliable abundances. We relax this threshold at low $\mgh$ to maintain an adequate sample size in this range when studying the lower metallicity populations. If we maintain the ${\rm SNR} > 200$ cut for the low metallicity population, we will be left with 329 stars. Relaxing the SNR criterion to ${\rm SNR} > 100$ increases the sample size to 536 stars. As we will see in Section~\ref{sec:survey-design}, the sampling noise can be the limiting factor to probe correlations; this prompted us to favor a more lenient cut in SNR for the low-metallicity stars, even with the expense of a higher photon noise. We verified that our more lenient SNR cut does not incur much additional photon noise (because the measurements are still systematic dominated, see \S\ref{sec:discussion_uncertainties}). For the low metallicity sample, a ${\rm SNR} > 200$ criterion leads to a median $\sigma_{\rm [Fe/H]} = 0.010$ and $\sigma_{\rm [Mg/H]} = 0.014$, whereas a ${\rm SNR} > 200$ criterion leads to a median $\sigma_{\rm [Fe/H]} = 0.011$ and $\sigma_{\rm [Mg/H]} = 0.015$. However, for the more abundant Solar metallicity stars, sampling noise is a minor problem. Therefore, we chose to refine our correlation measurements by limiting ourselves to the high SNR sample. Nonetheless, since our study mostly focuses on the Solar metallicity stellar populations, this choice is not critical.

Our criteria are similar to those used by \citet{Weinberg2019}, but here we use DR16 instead of DR14. Also, they used a somewhat narrower $\logg$ range ($\logg = 1-2$) and had no separate $\teff$ cut. The number of stars passing our cuts is $20{,}367$. The 15 elements that we use in this study are the $\alpha$-elements Mg, O, Si, S, Ca, the light odd-$Z$ elements Na, Al, and K, the iron-peak elements V, Cr, Mn, Fe, Co, Ni, and the ``iron-cliff'' element Cu\footnote{We adopt stellar parameters and elemental abundances from the ``named tags'' attributes from the DR16 catalog (i.e., fits.TEFF, fits.LOGG, fits.MG\_FE). This further leads to a more restricted and cleaner sample as some combinations are $\teff$ and metallicities are automatically excluded (priv.~comm.~H.~J\"onsson).}. Although APOGEE measures C and N, we do not use them here because their atmospheric abundances in giant branch stars are affected by internal mixing and do not reflect the stars' birth abundances. At least for our well-curated APOGEE sample, we checked that there is no apparent bias in [X/Fe] for all elemental abundance ratios adopted in this study for stars with different C and N values. We define $\xh$ abundances from the reported APOGEE measurements as $\xh = \xfe + \feh$. Some stars that pass our global flag cuts have flagged values of $\xfe$ for individual elements. To keep our analysis straightforward, we further restrict the sample to stars with unflagged values for all 15 abundances and $-1.5 < [X/{\rm H}] < 1.5$ for all elements, reducing it to $20{,}111$ stars.

In Fig.~\ref{fig4}, we illustrate the abundance distribution of our APOGEE training set in the $\xfe$ and ${\rm [Fe/H]}$ plane. All elemental abundances show a well-defined locus. There is larger dispersion for Na and V, and to some extent K and Cu, but the more considerable dispersion is not surprising as these elements only have weak or singular features in the APOGEE $H$-band spectra. The statistical uncertainties reported for these elements are also larger than for other elements.

To demonstrate the ability of the normalizing flow to emulate this distribution, we first train a normalizing flow with this training set and fit for the joint distribution of all 15 elemental abundances, $p(\xh)$. We then sample from this 15D joint distribution. The density contours in Fig.~\ref{fig5} demonstrate the sample of $10^6$ drawn from the fitted normalizing flow. Comparing Fig.~\ref{fig4} and Fig.~\ref{fig5} showcases the remarkable ability of the normalizing flow to represent the APOGEE abundance distribution. From here onward, unless stated otherwise, we will focus on the conditional distribution, i.e., the joint distribution of some elements conditioned on the values of $\teff, \logg$, and two or more elements.

%
%
%
%
%
%

\subsection{The baseline: conditioning on Fe and Mg}
\label{sec:apogee_baseline}

We will first describe the abundance distribution of 13 elements, training a normalizing flow to describe $p(\xh \,|\, \feh, \mgfe, \teff, \logg)$\footnote{Conditioning on $\feh$ and $\mgfe$ is equivalent to conditioning on $\feh$ and $\mgh$, since the value of $\mgfe$ at fixed $\feh$ and $\mgh$ is just $\mgfe=\mgh-\feh$.}. We include $\teff$ and $\logg$ as conditioning variables because a star's elemental spectral features depend on these atmospheric parameters as well as on the abundances themselves. Due to the spectral models’ imperfection, this often translates into different measurement systematics for different stars. Conditioning on them allows us to study the abundances differentially, pushing the measurement uncertainties to approach those due only to photon noise. Our normalizing flow models also allow us to choose different reference points in $\teff$ and $\logg$ to evaluate the dispersions and the correlation matrices. Comparing results at different reference points allows us to test whether they are affected by systematic uncertainties within the range of our sample. Small differences could arise in principle because stars of different $\logg$ and $\teff$ have different luminosity, can have different SNR (hence photon noises) and sample the disk differently. However, the fact that median abundance trends are nearly independent of location within the disk or bulge \citep{Weinberg2019,Griffith2021} suggests that any genuine trends with disk sampling would be small.

As a baseline model, we also condition on Fe and Mg, which serve as representative elements for two critical enrichment processes, core-collapse supernovae and Type Ia supernovae. These two elements provide informative diagnostics for the contribution of these two processes to a star's abundances because (a) they are well measured by APOGEE, (b) Mg is expected to come almost exclusively from core-collapse supernovae, and (c) Fe has a large additional contribution from SNIa. By conditioning on these two elements we remove two dimensions that are known to be important in the Milky Way abundances, allowing us to study the residuals in finer detail.

Before studying the residual correlations, we first examine the diagonal entries of the covariance matrix. Recall that normalizing flows allow us to draw samples from the conditional distribution $p(\xh|\teff,\logg,\feh,\mgfe)$, with which we can evaluate the dispersion by drawing samples ($10^5$ in our case) from the conditional distribution. In Fig.~\ref{fig6}, we show the dispersion of the conditional PDF, conditioning only on these two elements. The figure illustrates, given a star's $\feh, \mgfe, \teff,$ and $\logg$ measurements in APOGEE, how well the conditional mean abundance predicts the other elemental abundances. The blue, orange, and green lines show the results for different reference values of the conditioning variables. We estimate the finite sampling uncertainty in the dispersion by constructing 640 bootstrap realizations of our $20{,}111$ APOGEE stars and repeating our entire procedure, training the normalizing flow for the conditional PDF on each bootstrap realization of the data. Unless stated otherwise all results in this study adopt 640 bootstrap realizations to calculate the finite sampling uncertainty. Furthermore, since the normalizing flow training itself can be noisy, we train 60 normalizing flows without bootstrapping and take the median of the covariances of these realizations as our best estimates.

On the left, we illustrate the dispersion for stars with different $\teff$ and $\logg$, assuming Solar metallicity (by which we mean both $\feh=0$ and $\mgfe=0$). We evaluate the dispersion of a given element about the conditional mean, denoted $\sigma_{[X/{\rm H}]}$, as half the difference between the 16th- and 84th-percentile values in the marginal PDF. Elements are listed by increasing order of this dispersion (blue line) for the $\teff=4500\,{\rm K}$, $\logg = 2.1$ conditional PDF. The dashed line shows the mean value of the reported {\sc aspcap} $\xfe$ uncertainty for all stars in our sample. The total dispersion is a nearly monotonic function of this estimated photon noise, but it is consistently higher, implying, if the {\sc aspcap} noise estimates are accurate, that there is residual intrinsic dispersion in the abundances. If we estimate this intrinsic dispersion as the quadrature difference between the total dispersion and the photon noise, we find values of 0.01-0.02 dex for most elements (0.007 dex for O and Co). The inferred intrinsic dispersion is larger (0.035-0.05 dex) for K, V, and Na. While these elements could truly have larger intrinsic dispersion, they are also three of the elements that are most difficult to measure with APOGEE spectra, so we suspect that this difference is a consequence of observational dispersion in excess of the estimated noise.

If we define $\sigma_{[X/{\rm H}]}$ as the rms deviation about the conditional mean instead of using the difference of percentile values, we get dispersions (not shown) that are slightly higher (5-10\%) for the best measured elements on the left side of the plot, but 25-70\% higher for the elements with the largest dispersion (Cr, Cu, K, V, Na). The larger rms values for these elements are driven by outliers on the tails of the PDF.\footnote{Although not shown, we found that performing a 3$\sigma$ clipping on the sample drawn from the conditional normalizing flows does not qualitatively alter the results of the correlation matrices presented in this study. Thus, our primary results are not noticeably affected by the outliers.} These outlier values could be real and might be astrophyically interesting, but we suspect that they are primarily non-Gaussian observational errors because they occur for the abundances that are most difficult to measure in the first place. If we used the rms deviation to infer the intrinsic dispersion, we would get larger values (0.03-0.08 dex) for these elements.

Green and orange lines show the dispersion at two other choices of $\teff$ and $\logg$, corresponding to successively cooler and more luminous stars. The residual dispersion is similar to that found for our fiducial $\teff$ and $\logg$ point, demonstrating the robustness of our results, irrespect of the chosen stellar parameters. There are minor differences, but those could be due to different photon noise at different reference points. We find that residual dispersions at fixed Fe and Mg are only slightly larger even if we do not condition on $\teff$ and $\logg$, which indicates that our parameter range is already narrow enough to limit the contribution of differential systematics.  We nonetheless retain $\teff$, $\logg$ conditioning for our default analysis, since (as argued in \S\ref{sec:measuring}) these parameters could be correlated with abundances, which complicates the interpretation of the correlations, even if conditioning on them makes only a tiny difference to the residual dispersion.

The right panel shows the residual dispersion for different metallicity and $\alpha$-enhancement. We investigate three different reference points, representing the Solar metallicity population and the low-$\alpha$ and high-$\alpha$ branches at low metallicity ($\feh=-0.5$ with $\afe = 0.05$, 0.25). The current APOGEE disk star sample has too few low metallicity stars to reliably investigate the abundance PDF below $\feh = -0.5$. Fig.~\ref{fig6} demonstrates that the dispersion about the conditional mean is qualitatively similar for these different populations. Some elements show larger dispersion at low metallicity, but these are mostly elements with weak spectral features in the APOGEE $H$-band, so the larger dispersion could be a consequence of larger observational uncertainties at low metallicity.

Strictly speaking, our use of the percentile range rather than rms deviation to define $\sigma_{\xh}$ means that the $\sigma_{\xh}^2$ are technically not the diagonal elements of the abundance covariance matrix. However, culling the outliers with the percentile range probably constitutes a better comparison with the reported photon noise uncertainty from {\sc aspcap}. We will ignore this terminological distinction below and use the terms diagonal covariance and dispersion to refer to the dispersion estimated by this percentile method, which responds to the core of the distribution rather than the tails.

%
%
%
%
%
%

\subsection{Dispersion alone cannot (yet) detect independent elements}
\label{sec:apogee_dispersion}

Fig.~\ref{fig7} shows the residual dispersion of abundances after conditioning on two elements (the baseline case discussed previously), three elements (Fe, Mg, and O), or four elements (Fe, Mg, O, and Ni). We adopt the reference point $\teff = 4500\,$K and $\logg = 2.1$ throughout. The left panel shows Solar metallicity stars and the right panel shows lower metallicity, $\alpha$-enhanced stars. Uncertainties in the residual dispersions are estimated from bootstrap resampling as before.

No reduction in dispersion is detectable at a statistically significant level. This result is unsurprising in light of our discussion in \S\ref{sec:measuring_variance}. Even if an element has correlations with other elements as strong as $\rho=0.4$, conditioning on that element only reduces the dispersion by $\sim 8\%$ on average, which requires an effective sample of $\sim \mathcal{O}(100)-\mathcal{O}(1000)$ for a 2$\sigma$ detection, within the finite sampling fluctuations. While the high quality APOGEE sample adopted here has $20{,}000$ stars, the effective sample at a reference ${\rm [Fe/H]}$, $\mgfe$ and stellar parameters is $\lesssim 500$ (see the discussion on the effective sample size in \S\ref{sec:apogee_sampling}). Stacking the signals at multiple reference points only moderately reduces the sampling noise, unless one takes such a large range of $\feh$ and $\mgfe$ that the results become more difficult to interpret.  Using a large sample of lower SNR spectra would not improve the signal either; in this case, the correlations become weaker due to the larger observation dispersions (Eq.~\ref{eqn:rhojktot}), which in turn would require an even larger effective sample (Eq.~\ref{eqn:Nmin_variance}) to measure the reduction of dispersion at high significance. Thus, from Fig.~\ref{fig7} alone, we might erroneously conclude that Fe and Mg contain all of the information concealed in APOGEE abundances.

As we will show in the following section, the elemental abundance space has many more hidden dimensions which manifest themselves through the correlations, but these dimensions are simply not visible in dispersion with the current limited sample size of APOGEE.  With larger samples in the near future (e.g., with SDSS-V, 4MOST, Weave), measuring other independent dimensions through the reduction in dispersion should become possible, though it remains highly inefficient compared to measuring correlations directly. Such results would validate the theoretical arguments as laid out in \S\ref{sec:measuring}. Finally, although not shown, we also tested that conditioning on any other combination of elements instead of O and Ni does not change the results.

\begin{figure*}
    \centering
    \includegraphics[width=1.0\textwidth]{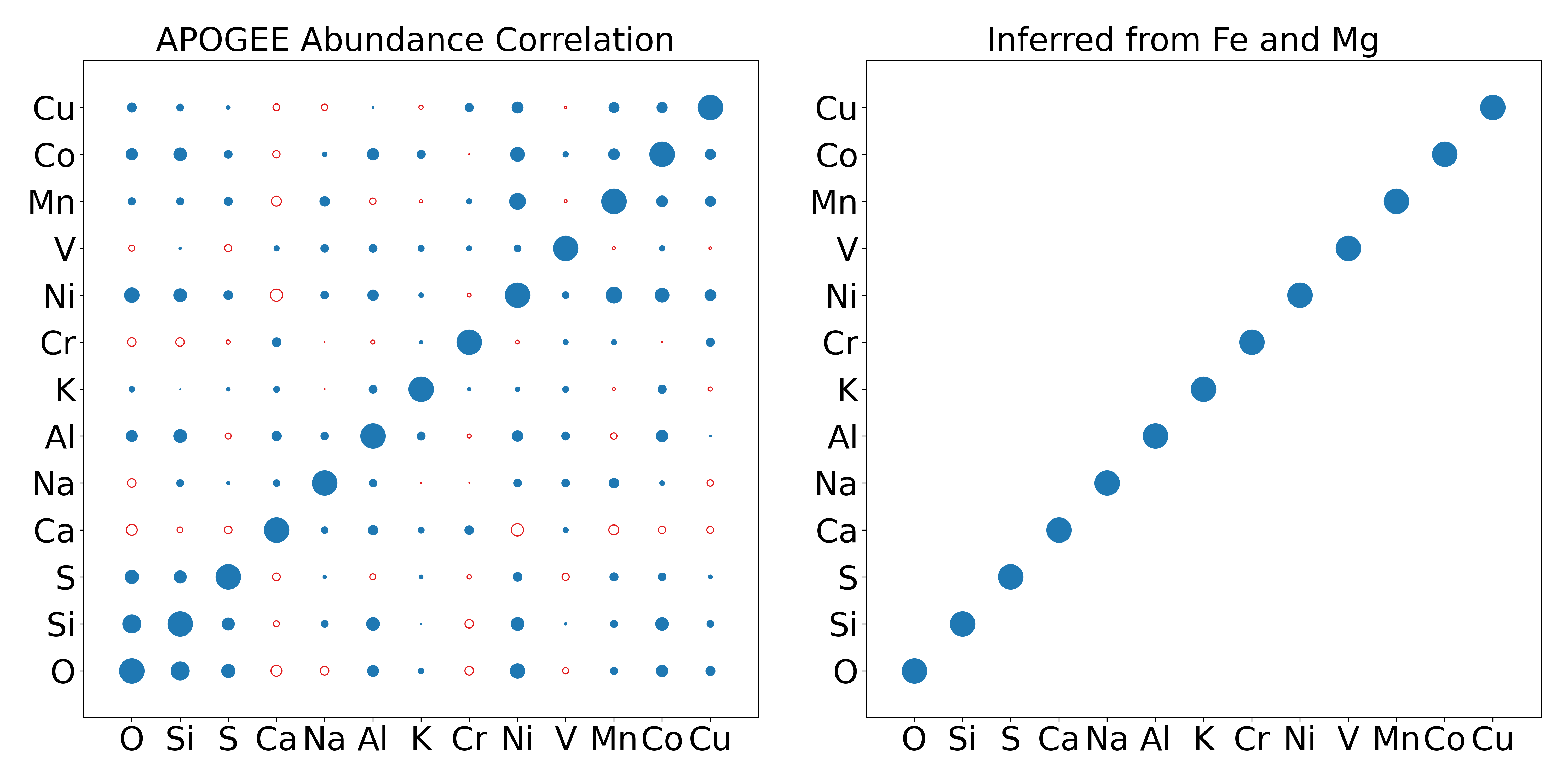}
    \caption{({\it Left}) Correlation matrix of 13 elemental abundances evaluated from the normalizing flow PDF conditioned on $\teff=4500\,{\rm K}$, $\logg = 2.1$, and Solar ${\rm [Fe/H]}$ and ${\rm [Mg/Fe]}$. Symbol areas are proportional to the magnitude of correlations, blue filled circles for positively correlated element pairs and red open circles for negatively correlated pairs. Diagonal entries of the correlation matrix have a value of 1.0 by definition. Even after removing the mean trends tracked by Fe and Mg, the APOGEE elemental abundance space has many ``hidden'' dimensions that only manifest themselves through statistical correlations of abundances. ({\it Right}) The trivial correlation matrix expected if we {\em infer} the 13 elemental abundances from the observed Fe and Mg abundances. Inferring abundances is not the same as measuring the abundances; only the latter reveals the information concealed in the subtle correlations, which we can now measure at high statistical significance from the extensive APOGEE data.}
    \label{fig8}
\end{figure*}

\begin{figure*}
    \centering
    \includegraphics[width=0.85\textwidth]{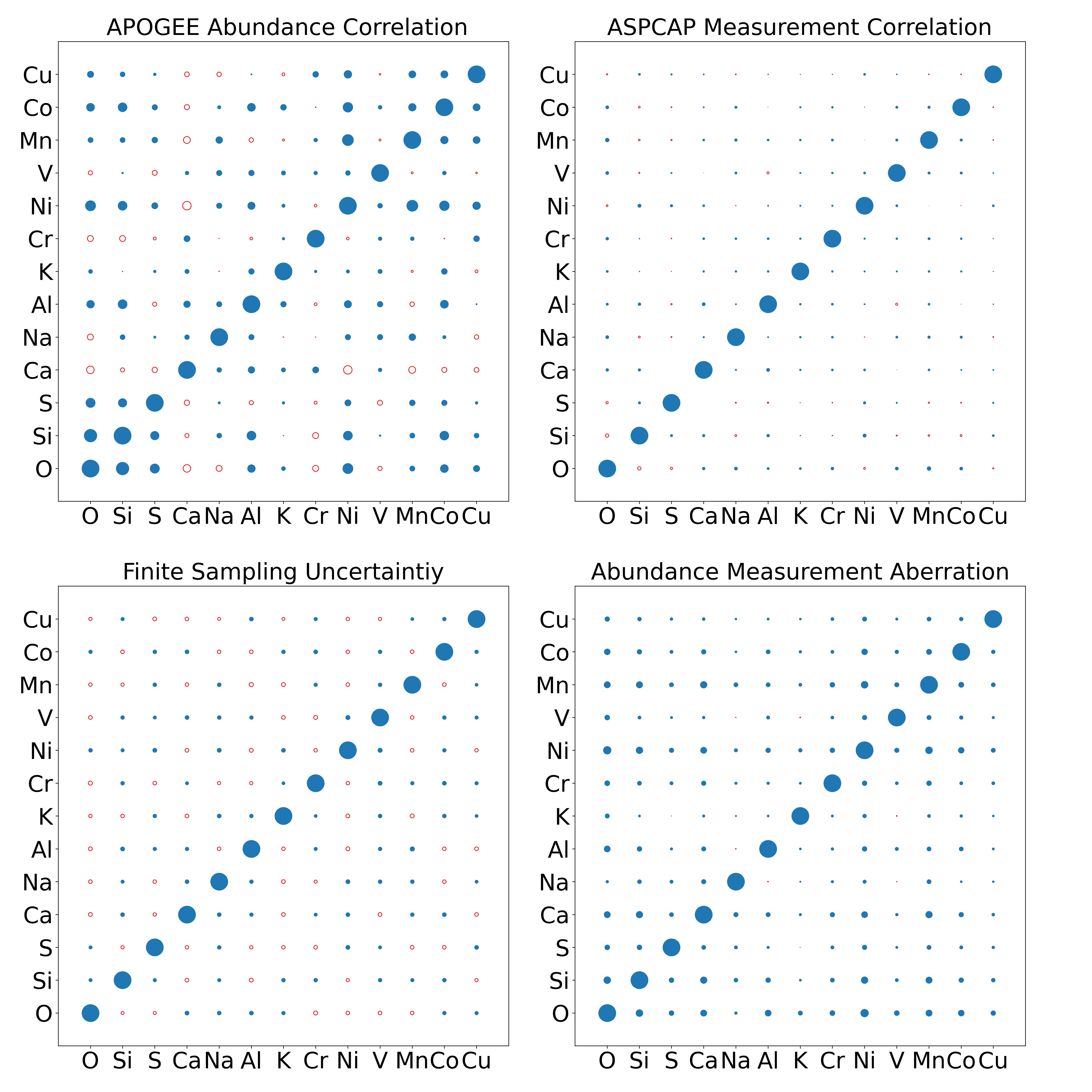}
    \caption{Comparison of the measured APOGEE correlations conditioned on Mg and Fe (the upper left panel), repeated from Fig.~\ref{fig8}, to three sources of observational uncertainty discussed in \S\ref{sec:apogee_aspcap}-\ref{sec:apogee_aberration}. The upper right panel shows the correlations expected from the impact of correlations due to the {\sc aspcap} measurements, which are not perfectly diagonal because values of some stellar parameters and blended features affect the inference of multiple elements. Off-diagonal entries in the bottom left panel show the magnitude of random uncertainties in correlation coefficients expected from finite sampling, estimated by bootstrap resampling of our APOGEE data set. Signs of these coefficients have been randomly chosen to emphasize that sampling uncertainty can be positive or negative. The bottom right panel shows the correlations expected from ``measurement aberration'' induced by conditioning on a star's measured values of ${\rm [Fe/H]}$ and ${\rm [Mg/H]}$ instead of the unknown true values. None of these sources of uncertainty is large enough to explain the correlation signals measured in APOGEE, indicating that the detection of the residual correlation structure in the data is statistically significant and astrophysically relevant.}
    \label{fig9}
\end{figure*}

%
%
%
%
%
%

\subsection{APOGEE data demonstrate residual correlation structure between elemental abundances}
\label{sec:apogee_correlations}

Besides dispersion, the sample drawn from the conditional distribution $p(\xh|\teff,\logg,\feh,\mgfe)$ also allows us to estimate the off-diagonal entries of the covariance matrix, and hence the correlation among the elemental abundances. The correlation matrix of the APOGEE data (assuming Solar metallicity, $\teff=4500\,$K, and $\logg =2.1$) is shown in the left panel of Fig.~\ref{fig8}. The figure shows that even after removing the mean abundance trends predicted by Fe and Mg, a non-trivial correlation between elements remains, implying higher dimensionality of the abundance distribution that would be missed if we considered only the residual dispersion. The right panel of Fig.~\ref{fig8} shows the trivial correlation matrix that we would obtain if other elemental abundances were perfectly determined by the observed Fe and Mg, leading to an identity correlation matrix. {\em Comparing the two panels of Fig.~\ref{fig8} makes the obvious point that inferring elemental abundances from Fe and Mg (right) is not the same as measuring them (left)}, even if they are statistically indistinguishable in terms of their dispersions (Fig.~\ref{fig7}). In \ref{sec:correlation-other-teff}, we further demonstrate that such correlation structure also shows up across different choices of $\teff$ and $\logg$, and is largely independent of the choice of stellar parameters.

However, a critical question remains: do the measured correlations reflect astrophysics, or could they be artificially induced by observational uncertainties? Three potential uncertainties could generate artificial correlations among elemental abundances; we will estimate each of these in turn and show that they are too small to explain the APOGEE signal.  Fig.~\ref{fig9} summarizes this comparison.

%
%
%
%
%
%

\subsubsection{Statistical covariances from the {\sc aspcap} measurements}
\label{sec:apogee_aspcap}

The first potential source of uncertainty is the correlated measurement uncertainty from {\sc aspcap}. For individual element abundances, {\sc aspcap} reports statistical uncertainties but it does not report the covariance of these uncertainties. Therefore, in the following, we provide our own estimate of the {\sc aspcap} correlation.

If we condition on $\teff$, $\logg$, ${\rm [Fe/H]}$, and ${\rm [Mg/H]}$, we expect to remove most differential systematic uncertainties as a source of dispersion or artificial correlations, and measurement uncertainties should approach the photon noise limit, as borne out in Fig.~\ref{fig6}. In this case, probing the {\sc aspcap} measurement covariance reduces to the question of understanding the Fisher matrix. For simplicity, we make the approximation that all pixels in the APOGEE spectrum have the same noise, and that the noise in different pixels is uncorrelated. We refer interested readers to \cite{Ting2017a} for details behind the calculations presented here. With these assumptions, it can be shown that the statistical covariance due to the photon noise (or what is known as the Cramer-Rao bound) is proportional to $(G \cdot G^T)^{-1}$, where $G$ is an $N_{\rm labels} \times N_{\rm pixel}$ matrix that collects all the gradient spectra. Each row in $G$ measures how an APOGEE spectrum would vary as we vary individual stellar labels (stellar parameters and elemental abundances).

To evaluate the gradient spectra, we adopt the Kurucz spectral models \citep{Kurucz1993, Kurucz2005, Kurucz2013} through {\sc atlas12/synthe} synthesizer. Since {\sc aspcap} measures abundances with spectral windows, we adopt the spectral windows from \citet{GarciaPerez2016} as well as additional spectral windows in DR16 for Cr, Co, and Cu (priv.~comm.~J.~Holtzman) and zero-out the gradient for any pixels that are not in the {\sc aspcap} window for the corresponding element. When calculating the statistical covariance matrix, besides the elemental abundances in this study, we also include gradient spectra from $\teff, \logg, v_{\rm micro}$, [C/H] and [N/H]\footnote{If we were to include also $v_{\rm macro}$ in the fit, it would have increased the median artificial correlation to $\rho \simeq 0.03$, instead of $\rho \simeq 0.01$, but the effect would still be negligible for this study.}. These are stellar labels that {\sc aspcap} also derived, and their measurement uncertainties could indirectly create artificial correlations among the elemental abundance uncertainties.

For ease of comparison, the top left panel of Fig.~\ref{fig9} repeats the measured APOGEE correlations shown previously in Fig.~\ref{fig8}. The top right panel shows the expected correlations from photon noise uncertainties, adopting the same reference point of $\teff =4500\,{\rm K}$, $\logg = 2.1$, and Solar metallicity. The covariance of photon noise abundance uncertainties for an individual star would be the product of these correlations with the individual element dispersions, and it would scale with the SNR of the spectrum. However, the correlation coefficients themselves are independent of the SNR. The figure shows that the correlations among elemental abundances expected from photon noise are minimal, with typical pairwise values $\rho \simeq 0.01$, much weaker than the empirical signals; the APOGEE correlation signals are of the order of $\rho = 0.2-0.4$. Our results echo those in figure 17 of \cite{Ting2017a}, who studied the correlations of abundance measurements at various resolutions and found that for the APOGEE resolution and wavelength coverage, most abundance measurements are uncorrelated even when blended features are included. Since {\sc aspcap} chose only to measure individual abundances through spectral windows without blended features, correlations between abundance measurements are even further reduced.

For completeness, we note that there are a few approximations that we have made for this calculation. For example, we adopt the Kurucz models instead of the {\sc marcs/turbospectrum} models adopted in {\sc aspcap}, as we do not have easy access to the latter. The difference in atomic data is likely to modify the derived element values slightly but have minimal influence on the correlations due to the photon noise. Similarly, we expect the assumption of uncorrelated pixels and homogeneous pixel noise might change the absolute scale of the covariance, but not the correlation by much. As we discuss in the following, the other two sources of correlated uncertainties are more important, and any artificial correlations due to the {\sc aspcap} measurements can be neglected for our purposes.

Another caveat is that empirical uncertainties from repeat spectra exceed those from $\chi^2$ fitting (see \S\ref{sec:measuring_uncertainties} and \citealt{Jonsson2020}), which implies that some variation in observational conditions (e.g., small changes in the spectral line spread function) contributes to statistical measurement uncertainties in addition to pure photon noise. Data reduction errors such as imperfect telluric subtraction or continuum determination could also produce correlated errors in principle, but in practice this is unlikely because abundances are determined from localized spectral features in wavelength regions that are largely disjoint for different elements.  We have done some simple experiments with idealized examples of such data reduction errors and find that they produce negligible correlations.  We henceforth assume any correlations arising from these additional random errors can be neglected.  This assumption could be tested empirically in the future by computing observational error covariances (and not just rms errors) using repeat spectra as in \cite{Jonsson2020}, but this approach requires significant changes in APOGEE data analysis procedures.

%
%
%
%
%
%

\subsubsection{Correlation estimation uncertainty from finite sampling}
\label{sec:apogee_sampling}

Another source of uncertainty that could generate artificial correlations is finite sampling. In the ideal scenario where we have infinite realizations drawn from the PDF, we should recover the PDF exactly. However, the finite sampling implies that the estimation of the conditional PDF itself, and subsequently the correlations, must be noisy to some extent. As derived in \S\ref{sec:measuring_covariance}, the uncertainty of correlation due to finite sampling is $\simeq 1/\sqrt{N_{\rm sample}}$. Quantitatively, we have the standard deviation of the correlation due to sampling uncertainty to be 0.1 for a sample size of 100 and 0.01 for a sample size of $10^4$.

Although we adopt a training set of $20{,}111$ stars in this study, not all stars contribute to any single reference point. Since we study the smooth variation of the conditional distribution and its correlation, it can be challenging to estimate the effective $N_{\rm sample}$ contributing to a given reference point. To do so we repeat our entire analysis procedure for 640 bootstrap resamplings of the full $20{,}111$ star data set and take the standard deviation of the derived correlation coefficients. The bottom left panel of Fig.~\ref{fig9} shows these sampling uncertainties, assuming the reference point at $\teff = 4500\,$K, $\logg = 2.1$, and Solar metallicity. The sign (positive or negative) of the fluctuation is randomly assigned to highlight that the sampling uncertainties can perturb the correlation estimates in either direction. The panel shows that the finite sampling uncertainties are typically $\rho \simeq 0.045$, small compared to many of the non-zero correlations that we measure from APOGEE. As a result, the measured APOGEE correlations cannot be entirely caused by the random fluctuations due to the finite size of the stellar sample.

Recalling that the statistical uncertainty is $\simeq 1/\sqrt{N_{\rm sample}}$ for large $N$ and weak correlations, we infer that the effective sample size at our chosen fiducial reference point is $N_{\rm sample} \simeq 500$. In principle, we could ``stack'' the correlation signals at different reference points to increase the effective sample. However, through numerical experiments we found that stacking the signals over different $\teff$-$\logg$ of our training sample ($\teff = 4100-4600\,$K) only reduces the sampling uncertainty slightly (from $\rho =0.045$ to $0.041$). The effective sample is much smaller than the parent sample due to the conditioning on ${\rm [Fe/H]}$ and ${\rm [Mg/Fe]}$, not $\teff$ and $\logg$. As we will see in \S\ref{sec:multiple-elements}, stellar populations with different metallicities and $\alpha$-enhancements exhibit subtle differences in the correlations. Therefore, although stacking the signals along ${\rm [Fe/H]}$ and ${\rm [Mg/Fe]}$ could in principle reduce the sampling uncertainty, it will come at the cost of interpretability. Moreover, at least for the case of conditioning on two elements, there is a larger source of uncertainty that we will discuss below. This uncertainty cannot be reduced with the effective sample size but rather depends on the abundances' measurement precision. Therefore, for simplicity and for keeping any residual systematic uncertainties under better control, we choose not to stack the results from different reference points.

%
%
%
%
%
%

\subsubsection{Abundance measurement aberration}
\label{sec:apogee_aberration}

The origin of the third source of uncertainties is more subtle, but it is the dominant source of artificial correlation for this study. Recall that, in the baseline model, we condition on Fe and Mg and study the residual covariances. However, even without any astrophysical correlation, the residual covariance will only approach the {\sc aspcap} measurement uncertainty plus sampling uncertainty if we condition on true abundance values of Fe and Mg. When we train the conditional normalizing flow, we can only condition on the {\em measured} values from APOGEE, not the true values; this limitation itself can generate some artificial correlations. For example, if we consider a set of elements that are strongly correlated with Fe, then in a star whose measured Fe abundance fluctuates low because of uncertainty, all of those elements will tend to appear high, in a correlated way, relative to the conditional mean. We refer to this effect as ``measurement aberration,'' by loose analogy to the phenomenon of aberration of starlight. It is an uncertainty that arises because we are ``standing in the wrong place,'' predicting a star's conditional mean abundances based on its measured abundances of Fe and Mg instead of their true values.\footnote{For analytic discussion see \S 8.2 of \cite{Weinberg2021}}.

We estimate this effect through numerical experiments. In particular, we adopt the empirical conditional distribution $p(\xh \, | \, \teff, \logg, \feh, \mgfe)$ and its corresponding covariance matrix as shown in Fig.~\ref{fig8}. We then draw a mock sample that has the same $\teff, \logg$, $\feh$, and $\mgfe$ values as the stars in our APOGEE training set. Instead of drawing $\xh$ from the joint distribution, we draw each element independently from its own marginal distribution, generating a test sample that follows the same empirical dispersion as the APOGEE data but without the correlation. The elemental abundance space spanned by the mock data is strictly two-dimensional by construction, as Fe and Mg determine all abundances without any residual correlation. To study the aberration effect, we then add observational uncertainty to $\feh$ and $\mgh$, assuming the mean {\sc aspcap} reported uncertainties for our sample, $\Delta {\rm [Fe/H]} = 0.008\,$ dex and $\Delta {\rm [Mg/H]} = 0.011\,$ dex. Since we draw the other 13 elemental abundances from their own marginal distributions, their observational dispersions as shown in Fig.~\ref{fig6} are already automatically included. We refit a new conditional normalizing flow and study the correlation of this mock sample. To minimize the sampling uncertainty in this aberration estimate (because we have to draw the conditioning variables from the APOGEE data set, which is finite), we run the experiment 60 times, each time drawing new randomly perturbed values of the conditioning variables from the APOGEE data. For individual correlation coefficients, we take the median of these 60 realizations as our best estimates.

The bottom right panel of Fig.~\ref{fig9} shows the artificial correlations due to this effect. As we will elaborate more later with Fig.~\ref{fig11}, this source of artificial correlations is not negligible. Even though the artificial correlations peak at $\rho = 0.05$, they have a long tail extending to $\rho = 0.1-0.2$. Nevertheless, this effect is not sufficient to explain the strongest APOGEE correlations ($\rho = 0.2-0.4$). Furthermore, the correlations induced by measurement aberration can be straightforwardly predicted by the numerical experiment conducted here under the ``null hypothesis'' that all abundances are determined by Fe and Mg. Deviations from this predicted structure are therefore evidence against the null hypothesis. As we condition on more elements (\S\ref{sec:multiple-elements}), the aberration effect changes and diminishes because the random uncertainty in any one abundance matters less, so we redo the aberration prediction for each new null hypothesis (see Fig.~\ref{fig11} below).

The main uncertainty in predicting the aberration effect is that we rely on the {\sc aspcap} value of the photon noise uncertainty. In \ref{sec:uncertainty_aberration}, we show that generating artificial correlations as strong as $\rho =0.2-0.4$, the largest values we find for the APOGEE data, would require that {\sc aspcap} has underestimated the statistical uncertainties for Fe and Mg by a factor of $\sim 2$, with $\Delta {\rm [Fe/H]} = \Delta {\rm [Mg/H]} = 0.02\,$dex. However, in this case the structure of the correlations would be radically different, with all elements positively correlated. This also goes against the fact that the observed total dispersions (including intrinsic dispersions) for some elements with even less spectral information in APOGEE, such as O and Si, are close to 0.01 dex (Fig.~\ref{fig6} and Fig.~\ref{fig7}). This contradiction is itself indirect evidence that {\sc aspcap} is indeed achieving {\em differential} metallicity precision at the 0.01 dex level for Mg and Fe (as well as O and Si), consistent with the reported photon noise uncertainties. The realization of such exquisite precision in a large survey with mass production ``pipeline abundances'' is a remarkable achievement.  While absolute or differential systematic uncertainties are a limiting factor for some investigations, the high numerical precision attained by APOGEE can be harnessed for many applications with proper statistical modeling.

To sum up, through an exhaustive search for false positive signals, we conclude that the observed APOGEE correlations are real and statistically significant. They cannot be explained away by measurement uncertainties.

\begin{figure*}
    \centering
    \includegraphics[width=1.0\textwidth]{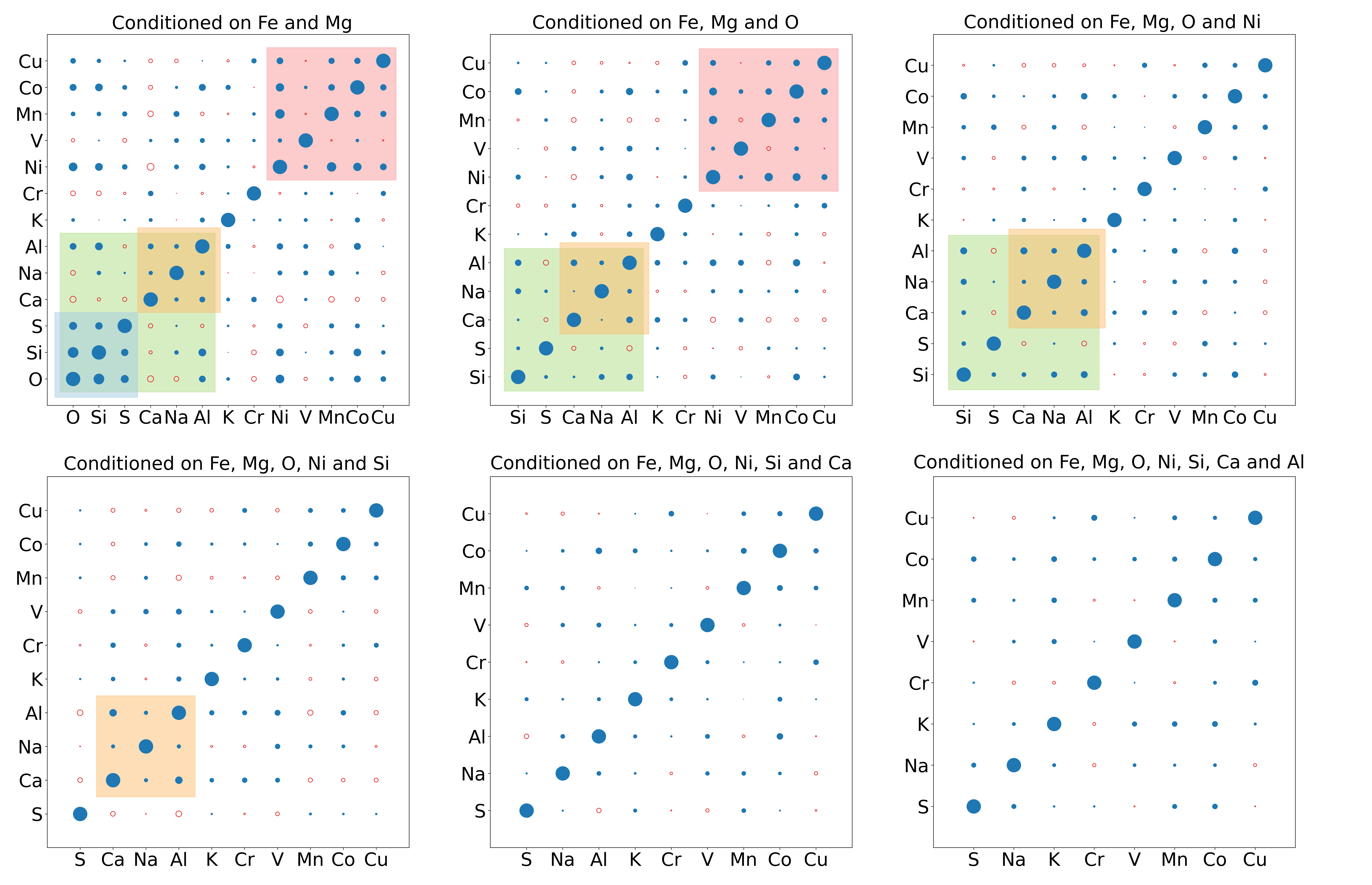}
    \caption{Correlation of residual abundances after conditioning on two, three, four, five, six, or seven elements as labeled. Shaded regions show blocks of correlations to guide the eye. Correlations within these blocks are reduced after conditioning on one of their constituent elements. All correlations are evaluated at $\teff=4500\,{\rm K}$, $\logg = 2.1$, and $\feh= [X/{\rm Fe}] =0$. The strong correlations present after conditioning on Fe and Mg alone (top left) are reduced to a level consistent with observational uncertainties by conditioning on Fe, Mg, O, Ni, Si, Ca, and Al (bottom center). See Fig.~\ref{fig11} for a more quantitative assessment.}
    \label{fig10}
\end{figure*}

\begin{figure*}
    \centering
    \includegraphics[width=0.75\textwidth]{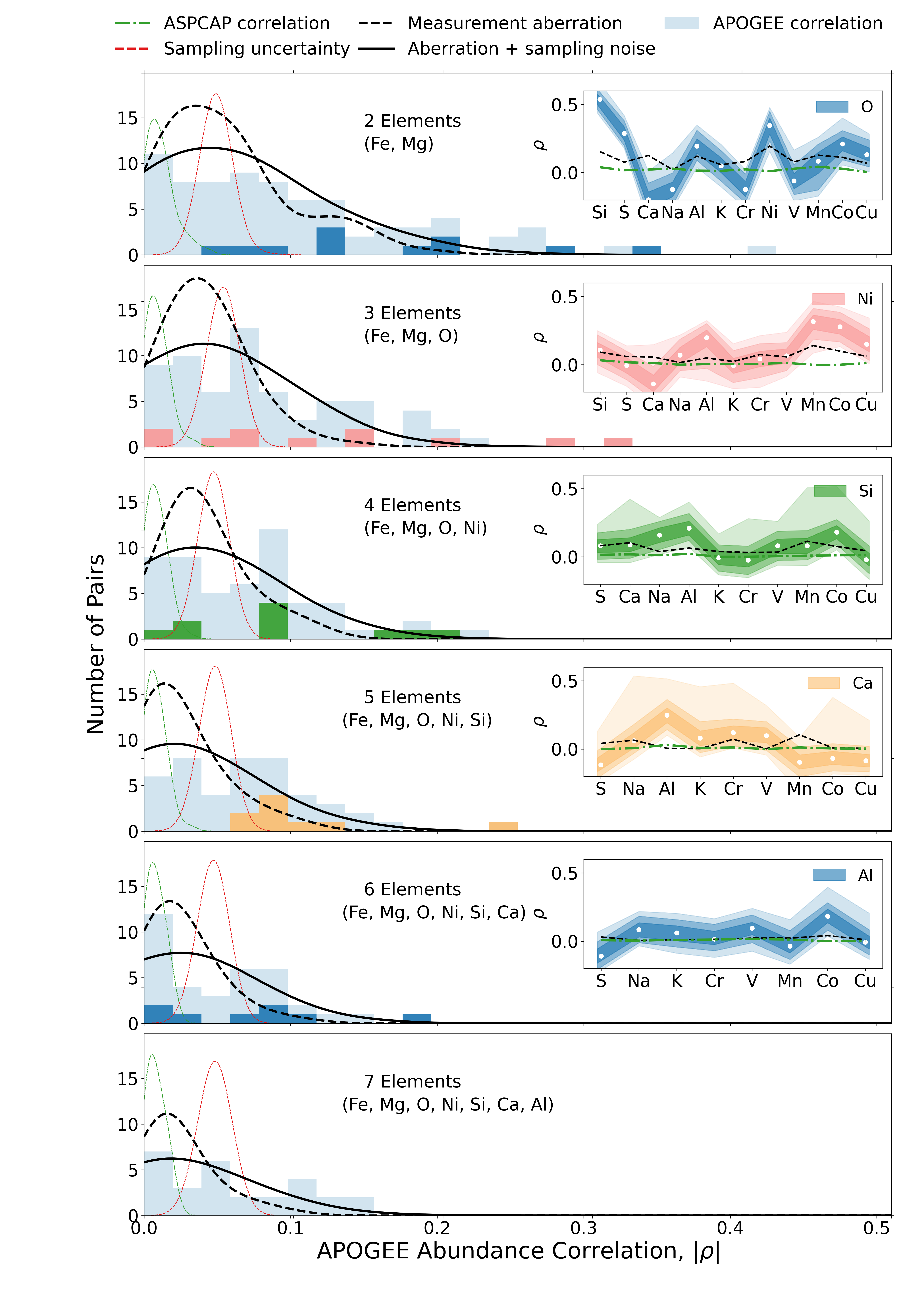}
    \vspace{-0.5cm}
    \caption{Statistical significance of residual correlations after conditioning on two to seven elements (top to bottom). In each panel light blue histograms show the distribution of $|\rho|$, the magnitudes of off-diagonal correlations in the corresponding panel of Fig.~\ref{fig10}. Green, red, and dashed black curves show the distribution of correlations expected from correlated observational uncertainties, finite sampling uncertainty, and measurement aberration, respectively. Solid black curves show the combined effect of sampling uncertainty and aberration. Inset panels show the correlation coefficients for the indicated element, with 68, 95, 99 percentile ranges estimated from the percentiles of 640 bootstrap resamplings of the APOGEE data set. For the first four panels, band colors are those of the corresponding element blocks in Fig.~\ref{fig10}, and these correlations are highlighted in the main panel histogram. Green dot-dashed and black dashed lines in the insets show the correlations predicted from statistical uncertainties and measurement aberration, respectively. Conditioning on seven elements is required to reduce the measured correlations to values that are all individually consistent at 95 percentile with the ``null hypothesis'' of no further residual correlations.}
    \label{fig11}
\end{figure*}

%
%
%
%
%
%

\subsection{Many APOGEE elemental abundances contain independent information}
\label{sec:multiple-elements}

After demonstrating that the APOGEE residual correlation structure is statistically significant, we turn to the question that we are the most interested in: How many APOGEE elements carry independent information? In other words, starting from the baseline model conditioned on Fe and Mg, which other elements we should condition on to reduce residual correlations to a level consistent with observational uncertainties? Because of the measurement uncertainties and finite sample size, our results will be a lower limit to the number of elements with intrinsically significant information content. Furthermore, observational uncertainty suppresses correlations (Eq.~\ref{eqn:rhojktot}), so our ability to detect correlations is reduced for the elements with the largest uncertainties.

Fig.~\ref{fig10} presents an overview of our principal results, which we will elaborate more quantitatively in Fig.~\ref{fig11}. Successive panels show the residual correlations after conditioning on two, three, four, five, six, or seven elemental abundances, always at the reference point $\teff = 4500\,$K, $\logg = 2.1$, $\feh=[X/{\rm Fe}] =0$. The elements are sorted by their commonly associated yield channels -- $\alpha$-elements (O, Si, S, and Ca, in addition to Mg), light odd-$Z$ elements (Na, Al, K), and the iron peak elements (Cr, Ni, V, Mn, Co, and Cu, in addition to Fe). Shaded blocks highlight groups of elements within a yield channel that show strong internal correlations, and each new conditioning element is chosen to target one of these blocks. The strong correlations among O, Si, and S in the top left panel are reduced by conditioning on O. Further conditioning on Ni reduces correlations among iron peak elements that remain after conditioning on Fe, Mg, and O. Conditioning on Si reduces several remaining correlations among $\alpha$-elements and the light odd-$Z$ elements Na and Al. Significant correlation remain among Ca, and Al, which is reduced by conditioning on Ca. Finally, although it is hard to see from Fig.~\ref{fig10}, there is a statistically significant (Fig.~\ref{fig11}) anti-correlation between S and Al, which is reduced by conditioning on Al. We note that the shaded blocks are only meant to guide the eyes. Even if a single yield channel is responsible for producing all elements within the same group, we might not expect to see strong internal correlations because of two confounding factors: (a) elements with large uncertainties (e.g., V) will show weaker observed correlation (see Section~\ref{sec:measuring_covariance}), and (b) there are inter-group correlations that make disentangling the correlations non-trivial.  We discuss interpretation of these correlations in \S\ref{sec:discussion_nucleosynthesis}.

Fig.~\ref{fig11} shows the statistical significance of these correlations. In the top panel, the light blue histogram shows the distribution of the magnitudes of the correlation coefficients after conditioning on Fe and Mg, i.e., of the off-diagonal elements of the matrix in the top left of Fig.~\ref{fig10}. The dark blue histogram shows the correlation coefficients involving O, which include several of the largest values in the distribution. In the inset panel, the band shows the O correlations element by element, with finite sampling uncertainties computed from the 68, 95, and 99 percentile range of the 640 bootstrap resamplings of the data set (\S\ref{sec:apogee_sampling}). The green dot-dashed line shows our estimate of the correlations from photon noise in the {\sc aspcap} abundance measurements (\S\ref{sec:apogee_aspcap}), which are small enough that we can neglect them relative to other sources of correlation. The black dashed line shows the correlations expected from measurement aberration (bottom right panel of Fig.~\ref{fig9}), computed as described in \S\ref{sec:apogee_aberration}. This line represents the prediction of the ``null hypothesis'', computed from 60 realizations in which we add random uncertainty to ${\rm [Fe/H]}$ and ${\rm [Mg/H]}$ values in a model that determines all abundances from Fe and Mg by construction (\S\ref{sec:apogee_aberration}). The O-Si correlation is highly inconsistent with this null hypothesis, and the O-S, O-Ca, and O-Cr correlations are all inconsistent at well over the 99 percentile range. Although not shown, we also tested that even if we were to include uncertainty range from the estimate of the aberration (from the 60 independent numerical experiments), instead of just taking the median prediction for our null hypothesis, the detection signals are still over the 99 percentile range.

Returning to the main panel, the green dot-dashed and red dashed curves show the distribution of correlation coefficients from {\sc aspcap} uncertainties (Fig.~\ref{fig9}, top right) and finite sampling (Fig.~\ref{fig9}, bottom left), respectively. The black dashed curve shows the distribution of correlations induced by measurement aberration. The solid black curve shows the combined effect of measurement aberration and sampling uncertainty, obtained by adding random draws from the sampling uncertainty distribution for a given coefficient to the median measurement aberration for the same coefficient. Many of the correlations measured from the APOGEE data are well beyond the tail of the distribution expected from measurement aberration and sampling uncertainty alone.

The second row shows the same quantities after conditioning on Fe, Mg, and O. Now the inset panel shows correlation coefficients for Ni, which has several of the largest values. Both Ni-Co deviate from the measurement aberration prediction over the 99 percentile range, and Ni-Mn, Ni-Ca are at the 95-99 percentile level. We emphasize that the sampling uncertainty and measurement aberration must be recomputed each time a new conditioning element is added. The measurement aberration effect gets gradually smaller as more conditioning elements are included because the random uncertainty in any one abundance measurement has less impact and is less likely to generate correlated aberration. The sampling uncertainty distribution changes slightly because the uncertainty for individual coefficients depends on the strength of the correlation. The {\sc aspcap} uncertainty matrix does not need to be recomputed as more conditioning elements are included, but rows or columns including those elements are omitted.

After adding Ni as a conditioning element, the Si-Al, Si-Na, and Si-Co correlations show the most significant deviations from the measurement aberration prediction (fourth row). The first Si-Al correlation is over the 99 percentile range, and Si-Na and Si-Co are at the 95-99 percentile range. After adding Si as well, the most significant deviations are Ca-Al and Ca-Mn (fifth row), both 99 percentile range. In this five-element case, the overall distribution of $|\rho|$ is consistent with the combination of measurement aberration and sampling uncertainty (main panel), but the specific Ca-Al and Ca-Mn correlations are not (inset) because the measurement aberration value for the Ca-Al coefficient is small, and the value for Ca-Mn is opposite in sign from the observed correlation. Finally, in the six-element case, the Al-S (95-99 percentile) and Al-Co (99 percentile) correlations remain significant, which we further reduce by conditioning on Al.

We emphasize, even when only conditioning two elements, we are left with $13 \times 12 / 2 = 78$ pairs of elemental abundances, and we typically expect $78 \times 1\% < 1$ pair to show correlations beyond 99 percentile. Therefore, {\it any} pair showing 99 percentile correlation is statistically significant by itself. More importantly, for most elements, we have multiple pairs that show significant correlations, which makes the combination of {\it all} these correlations appearing by chance highly unlikely. Nonetheless, since {\it individual} correlation signals are still at the $\sim 2.5-3\sigma$ level, a larger spectroscopic sample in the future will be critical to confirm and further validate these results, beyond what we can achieve with the current APOGEE sample.

After adding Al as a seventh conditioning element, the largest residual correlations are all consistent with measurement aberration + sampling uncertainty at the 95 percentile level (seventh row). We therefore do not claim convincing evidence of residual correlations beyond seven elements.

There is some judgment involved in deciding the order in which to add conditioning elements. Here we have made these choices based on both the magnitude of the residual correlations and the statistical and systematic uncertainties in the abundance measurements, skipping over some elements for which APOGEE measurements are less robust (e.g., Na). We have checked that alternative orderings lead to the same conclusion about the number of elements required to reduce residual correlations to a level consistent with observational uncertainty, though the choice and order of {\it which} seven elements to condition on is not unique. The elements that most clearly demonstrate residual correlations are also the seven with among the smallest {\sc aspcap} measurement uncertainties and the smallest total dispersion (see Fig.~\ref{fig6}). We suspect that improving the photon noise uncertainty of the abundance measurements would show that even more elements contain significant independent information.

\begin{figure*}
    \centering
    \includegraphics[width=1.0\textwidth]{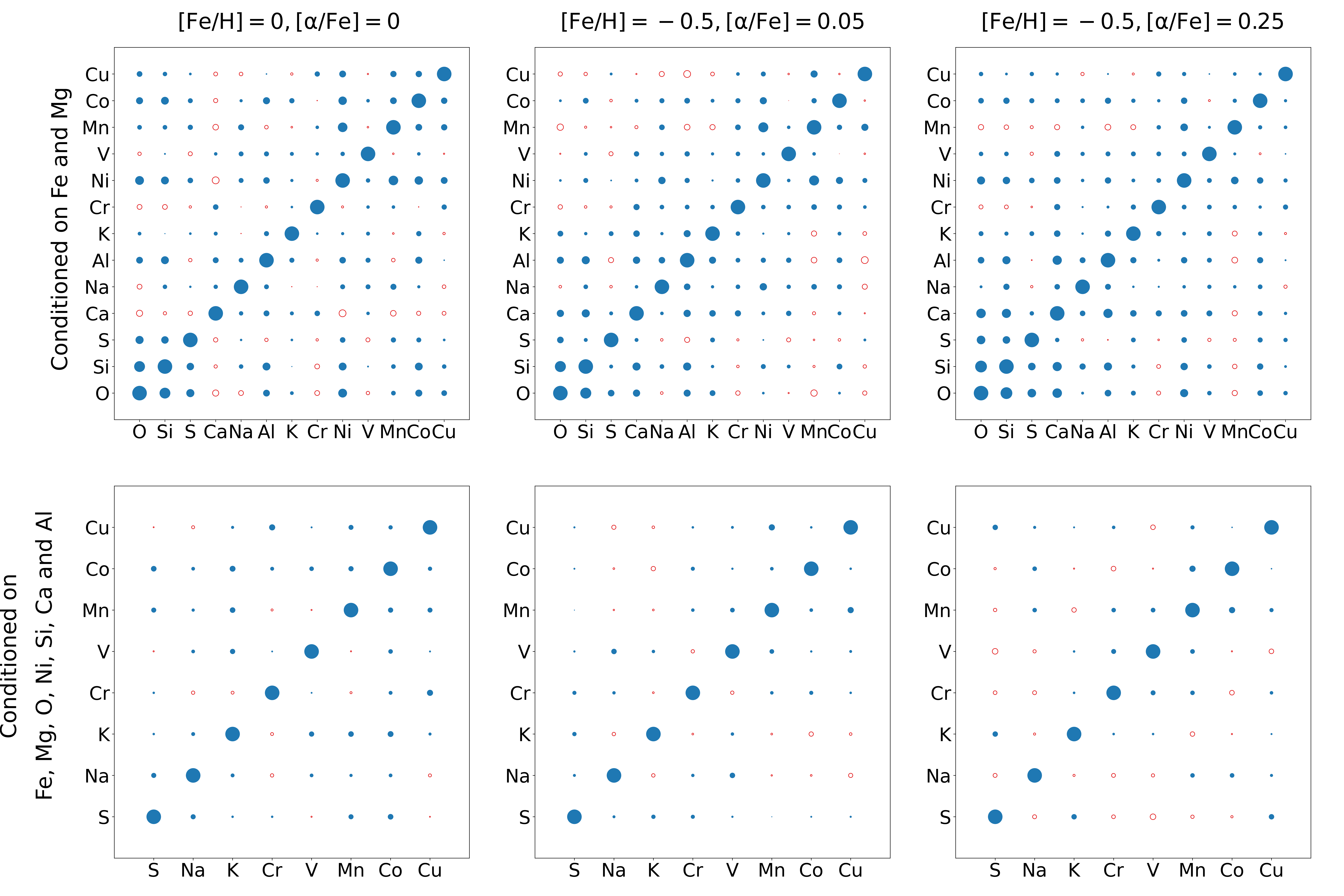}
    \caption{Correlation matrices for Solar metallicity stars (left) compared to those of $\feh=-0.5$ stars with $\mgfe=0.05$ (middle) or $\mgfe=0.25$ (right). Top and bottom rows show conditioning on two and seven elements, respectively. Residual correlation patterns in the top row are similar but not identical for different populations, and they are reduced to a similar level by seven-element conditioning as shown in the bottom panels.}
    \label{fig12}
\end{figure*}

Finally, stars with different ${\rm [Fe/H]}$ and ${\rm [Mg/Fe]}$ sample stellar populations that have experienced different enrichment histories and potentially different degrees of stochasticity in their chemical evolution. Fig.~\ref{fig12} compares the residual correlations for Solar metallicity stars (left) to those for $\feh=-0.5$ and $\mgfe=0.05$ (middle) or 0.25 (right), always with $\teff = 4500\,{\rm K}$ and $\logg = 2.1$. The residual correlations for metal-poor stars are comparable in magnitude and similar in pattern to those for Solar metallicity stars, but with some differences. Most noticeably, correlations involving Ca are stronger and consistently positive for the metal-poor stars. For the $\alpha$-enhanced stars the correlations among the $\alpha$-elements are somewhat stronger and those among the iron peak elements somewhat weaker. These differences are not surprising given the greater relative contribution of core-collapse supernova enrichment to the high-$\alpha$ population, though we caution that the residual correlations after conditioning on Fe and Mg need not follow the average contribution of individual enrichment processes in a simple way (see \S\ref{sec:discussion_dimensionality}). Importantly, as for Solar metallicity stars, the residual correlations reveal structure in the abundance distributions that would be buried if we were to study only the dispersion (Fig.~\ref{fig7}). The bottom panels show that conditioning on seven elements again removes most of the large correlations, though we have not investigated the significance of correlations as exhaustively for these low metallicity populations.

%
%
%
%
%
%

\section{Discussion}
\label{sec:discussion}

The analysis in \S\ref{sec:multiple-elements} shows that one must consider at least seven elements (Fe, Mg, O, Si, Ca, Ni, and Al) to remove residual correlations in the conditional PDF of APOGEE abundances. These elements are also among the most precisely measured in APOGEE data, and they display the smallest total dispersion after conditioning on Fe and Mg (all but Al have dispersions $<0.02$ dex, and Al has a dispersion of 0.027 dex, see Fig.~\ref{fig6}). With numerical experiments, we found that if we were to add 0.03 dex of noise, most measured correlations as shown in Fig.~\ref{fig11} would be consistent with just the photon noise and the measurement aberration prediction. When adding 0.04 dex of noise, most correlations are dominated by aberration. Therefore, the dimensionality is most likely limited by how many elemental abundances we can measure at the level of 0.02-0.03 dex, and we suspect that most or all elemental abundances would show significant residual correlation structures in data with still higher measurement precision. Crucially, these correlations can only be discovered if abundances are measured individually, not inferred based on the abundance of other elements.

Our correlation measurements are made possible by a new statistical technique, powered by the latest technology in machine learning, to model the high-dimensional and irregular distribution of stars in elemental abundance space. The technique allows us to mitigate systematic uncertainties by conditioning on stellar parameters ($\teff$, $\logg$) that affect abundance measurements. Further conditioning on Fe and Mg reduces residual dispersion about the conditional mean to 0.01-0.04 dex for most elements, and it reveals the sub-0.02-dex intrinsic dispersion after subtracting the reported {\sc aspcap} statistical uncertainties in quadrature. Our discussion in \S\ref{sec:measuring} explains why detecting hidden dimensions through reduction of residual dispersion is statistically difficult, and our analysis in \S\ref{sec:apogee} bears out this expectation in practice.

Instead, these complex but critical signatures of chemical enrichment processes can be revealed by directly measuring cross-element correlations in conditional PDFs. Conditioning on Fe and Mg alone leaves residual correlations that clearly exceed the levels expected from correlated measurement uncertainties, from statistical fluctuations due to finite sample size, or from the measurement aberration caused by random uncertainties in the conditioning abundances. We need to condition on Fe, Mg, O, Ni, Si, Ca, and Al to reduce the residual correlations for Solar metallicity stars to a level that is arguably consistent with the observational uncertainties.

In this section we discuss the implications of our results for the dimensionality of the stellar distribution in elemental abundance space, for characterization of observational uncertainties in abundance measurements, for chemical tagging, and for design of Galactic spectroscopic surveys.

%
%
%
%
%
%

\subsection{Characterizing dimensionality and the stochastic chemical evolution of the Milky Way}
\label{sec:discussion_dimensionality}

Our findings help to resolve the tension identified in \S\ref{sec:intro} between studies showing that disk star abundances can be accurately predicted from $\feh$ and age \citep{Ness2019} or from $\mgh$ and $\mgfe$ \citep{Weinberg2019} and PCA analyses showing that 5-10 principal components are required to explain the diversity of stellar abundance patterns \citep{Ting2012,Andrews2012} or APOGEE spectra \citep{Price-Jones2018}. Conditioning on Fe and Mg does indeed reduce residual dispersion to a level that only moderately exceeds that expected from photon noise. However, cross-element correlations clearly demonstrate the presence of underlying residual structure in the abundance patterns beyond the star-by-star dispersion.

The question of how many elements are needed to remove residual correlations is closely connected to the more general question of the dimensionality of the stellar distribution in elemental abundance space. If we have measurements of $M$ abundances for every star, then these measurements define an $M$-dimensional space, but the stars may lie along a one-dimensional curve, a two-dimensional surface, a three-dimensional hypersurface, etc. We expect the number of dimensions to be connected to the number of distinct astrophysical processes that contribute to the elements being considered. However, the connection is indirect. For example, in a one-zone model with a {\em fully mixed} interstellar medium (ISM), stellar abundances depend on a single parameter (time), even if the star formation history is complex and there are many processes contributing to the elements (core-collapse supernovae, Type Ia supernovae, AGB stars, neutron star mergers, etc.). While the relative contribution of these processes changes with time, all stars of the same age have the same integrated contributions.

Adding dimensionality to the abundance distribution thus requires mixing stellar populations that have experienced different enrichment histories. Radial migration of stars is one such mixing mechanism: star formation, accretion, and outflow histories within the Galactic disk change systematically with radius, and the stars present at a given radius today were born in a range of annuli with a variety of chemical evolution tracks \citep[e.g.,][]{Schoenrich2009,Minchev2013,Frankel2020}. Incomplete mixing of the ISM, in azimuth at a given radius or even within a single star-forming complex, allows stars to be born with a variety of abundance patterns at nearly the same location and time \citep{Krumholz2018}. Bursts of star formation produce sharp excursions of $\afe$ ratios and complex evolutionary tracks for elements produced by AGB stars on a timescale separate from that of core-collapse supernovae or Type Ia supernovae \citep[e.g.,][]{Johnson2020}. Mergers are another mechanism for mixing stellar populations with different histories, though the thinness of the Galactic disk implies that the fraction of disk stars that originated in a distinct satellite should be small \citep[e.g.,][]{Toth1992,Ting2019}.

Our analysis here implies a lower limit of seven for the effective dimensionality of the APOGEE disk abundance distribution. We note that this number is clearly a conservative limit. The elements studied in this paper are expected to come predominantly from core-collapse and Type Ia supernovae (see \citealt{Andrews2017,Rybizki2017,Weinberg2019}).  But even within the most ``boring'' elements of disk stars at Solar metallicity, our study reveals a plethora of abundance structures. An analysis including elements produced by other processes (e.g., optical surveys like GALAH or Gaia-ESO) will undoubtedly exhibit an even richer structure.

%
%
%
%
%
%

\subsection{Nucleosynthesis Implications}
\label{sec:discussion_nucleosynthesis}

Unfortunately, interpreting the residual correlations of Figs.~\ref{fig10} and~\ref{fig11} in terms of stellar enrichment sources remains challenging for several reasons.  First, while many of the correlation coefficients are convincingly larger than expected from measurement aberration and statistical uncertainties, they are not so much larger as to provide precise measurements of their values.  Second, the most obvious statistical tool to apply to multi-element abundance correlations, Principal Component Analysis \citep{Ting2012,Andrews2012,Andrews2017}, implicitly assumes that the latent factors governing these correlations are orthogonal.  There is no reason for the physical variations in stellar yield contributions to be statistically orthogonal, so the components recovered from PCA are often combinations of different stellar yields (as demonstrated by \citealt{Ting2012}).  Third, as already discussed in \S\ref{sec:discussion_dimensionality}, the residuals and their correlations depend on both the variations in stellar yield inputs and the processes that mix these variable yields in the ISM and that mix stellar populations within the disk.  Reliable interpretation of these correlation measurements will therefore require an ambitious forward modeling program that incorporates these physical processes and accounts for the impact of measurement uncertainties, perhaps using improved measurements from larger future data samples.

Here we offer some general interpretive comments, and we refer the reader to Weinberg et al. (\citeyear{Weinberg2021}; especially their \S 8) for more extensive discussion.  Instead of conditioning on Mg and Fe, \cite{Weinberg2021} fit each star's abundances with an empirical two-process model intended to capture core collapse and Type Ia supernova contributions, but because Mg and Fe provide precisely measured abundances that can distinguish these contributions the approaches are fairly similar in practice.  Despite the many differences of detail, including the use of APOGEE DR17 abundances rather than DR16, the correlations of abundance residuals found by Weinberg et al.\ (\citeyear{Weinberg2021}; see their Fig.~16) are qualitatively similar to those shown here in Fig.~\ref{fig10}.  In particular, both analyses find strong residual correlations among the $\alpha$-elements O, Si, and S, among the iron-peak elements Ni, V, Mn, and Co, and among the intermediate mass elements Ca, Na, and Al.\footnote{\cite{Weinberg2021} do not include Cu, which is not provided in the DR17 ASPCAP abundances, though they do include Ce (not available in DR16) and the element combination C+N.  They find that the correlations of Ca, Na, and Al extend to K and Cr, which is less obvious in our analysis.}

It is natural to associate the first of these correlated residual groups with variations in the core collapse supernova enrichment.  Stochastic sampling of the progenitor mass distribution might be sufficient to produce such variations depending on how efficiently ejecta are mixed within the ISM; this is a quantitative question that merits theoretical investigation.  Core collapse supernovae and Type Ia supernovae both contribute to the iron-peak elements, so variable enrichment from either population could drive correlated deviations in the second element group.  Type Ia supernovae may represent a heterogeneous population, e.g., with sub-Chandrasekhar and Chandrasekhar-mass progenitors, with single-degenerate and double-degenerate progenitors, with progenitors of different chemical composition, or with different explosion physics.  Variations in the relative contributions of populations with distinctive yields could produce variations among the iron-peak abundance ratios like those found here.

There is no obvious link among the elements in the intermediate group.  Theoretical models predict that core collapse supernovae dominate the production of Na \citep{Andrews2017,Rybizki2017}, but GALAH and APOGEE data show a substantial offset of [Na/Mg] between low-$\alpha$ and high-$\alpha$ stars implying a large contribution from a time-delayed enrichment mechanism \citep{Griffith2019,Weinberg2019,Weinberg2021}.  Possibly this source also contributes to Ca and Al, and variations in its amplitude lead to correlated deviations in these abundances at fixed Mg and Fe.  \cite{Weinberg2021} find a strong correlation between residual abundances of Na and of Ce, which is an $s$-process element expected to have a large contribution from AGB stars.  However, Al is expected to arise predominantly from core collapse supernovae \citep{Andrews2017,Rybizki2017}, and in contrast to Na it does not show distinct [X/Mg] ratios between low-$\alpha$ and high-$\alpha$ stars, which implies that its production mechanism is prompt like that of Mg.

These examples, especially the third, highlight the need for chemical evolution models that can make meaningful quantitative predictions for correlated abundance residuals at fixed Mg and Fe, and for the desirability of improved survey and analysis strategies (see \S\ref{sec:survey-design}) that can measure these correlations unambiguously and precisely.  It is only the size, homogeneity, and abundance precision of APOGEE that enables an empirical analysis like the one developed here, and the theoretical methods for exploiting such an analysis have yet to catch up.

%
%
%
%
%
%

\subsection{Characterizing observational uncertainties}
\label{sec:discussion_uncertainties}

In \S\ref{sec:measuring_uncertainties} we emphasized the distinction between ``absolute'' abundance uncertainties that are the same for all stars in a sample, differential systematic uncertainties that depend on stellar parameters such as $\teff$ and $\logg$ that vary across the sample, and statistical abundance uncertainties from photon noise. The first may dominate the difference between stars' measured and true abundances, but it does not contribute dispersion to abundances. Differential systematics can contribute dispersion, but they can be mitigated by conditioning on stellar parameters, a powerful feature of the normalizing flow method.

\begin{figure*}
  \centering
  \includegraphics[width=1.0\textwidth]{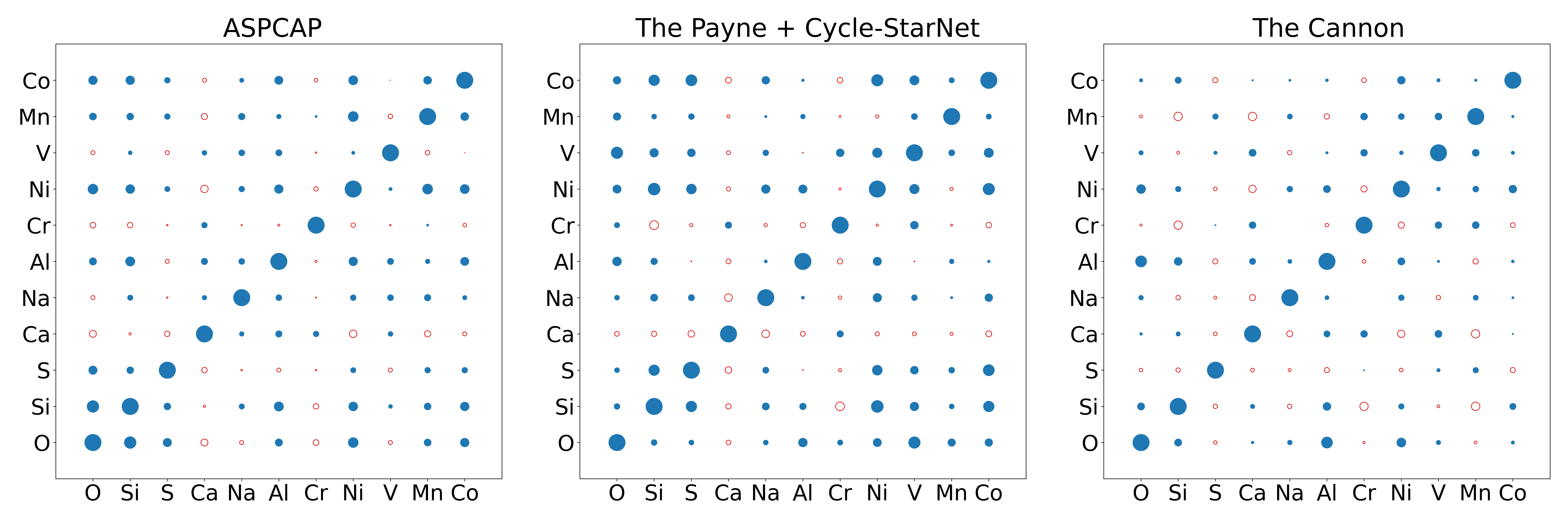}
  \caption{Residual correlation structures derived with different spectral analysis pipelines, all evaluated at our standard reference point $\teff=4500\,{\rm K}$, $\logg=2.1$, $\feh=\mgfe=0$. The left panel uses the {\sc aspcap} abundances (this study). The middle panel uses abundances for the same data set derived by {\sc the payne} with atmospheric models improved through {\sc cycle-starnet}. The right panel uses a DR14 sample with abundances from {\sc the cannon}. In all three cases, we adopt the common subset of 13{,}207 stars that have {\sc the cannon} DR14 abundances as the training set for a more robust comparison. The two {\it ab initio} modeling pipelines yield similar though not identical correlation patterns, while correlations from {\sc the cannon} are weaker on average and different in structure for some elements. Weaker correlations could arise if {\sc the cannon}'s abundances are partly affected by the fact that data-driven models partially infer abundances through astrophysical correlations instead of measuring them (see Fig.~\ref{fig8}).}
  \label{fig13}
\end{figure*}

By mitigating any differential systematics, we demonstrated that (Fig.~\ref{fig6}) the small statistical uncertainties reported by the {\sc aspcap} pipeline are an accurate representation of photon noise, with total dispersion including an intrinsic contribution of $0.01-0.02$ dex for well measured elements. We have also seen that even photon noise that is uncorrelated from pixel to pixel can cause correlated abundance errors because the abundances of multiple elements may be affected by the same uncertain stellar parameter, and because some abundances are estimated from blended features or molecular lines\footnote{Another source of correlation comes from the fact that some elements (e.g., essential electron donors) can substantially alter the stellar atmosphere. As a result, it would modify other elements' spectral features even those spectral features are not associated with the elements in question \citep[see the appendix in][]{Ting2016b} }. We have estimated these correlated measurement uncertainties for {\sc aspcap} in \S\ref{sec:apogee_aspcap} and find that they are small compared to statistical uncertainties in correlation coefficients from finite sampling (\S\ref{sec:apogee_sampling}). The largest source of artificial correlations is the ``measurement aberration'' arising because we can only condition on measured values of abundances rather than true values (\S\ref{sec:apogee_aberration}).

Although the photon noise alone might have limited impact on our ability to detect correlations at high significance, characterizing it is still enormously important. Fundamentally, we do not have direct access to the intrinsic variance and correlations, only to measured values that include observational contributions. Interpreting these measurements and tracing them back to astrophysical phenomena (\S\ref{sec:discussion_dimensionality}) requires a robust characterization of the intrinsic correlations, and hence an accurate determination of the observational uncertainties. While smaller statistical uncertainties are preferable, the most important thing is to understand them well enough so that their effects (including that of measurement aberration) can be removed.

This challenging goal is within reach as we approach a more {\it ab initio} way to perform full-spectral modelling. Full-spectral fitting is sometimes presented as a way to extract more information from blended features. Such an argument can be misleading because for high-resolution spectra this gain is minimal, as the information per spectral feature only adds in quadrature \citep{Ting2017a,Sandford2020}. However, in terms of extracting intrinsic covariance, performing full spectral fitting with all stellar labels simultaneously (e.g., with {\sc the payne}, \citealt{Ting2019b}, or {\sc cycle-starnet}, \citealt{OBriain2021}) may be advantageous compared to a multi-step approach like {\sc aspcap}'s; the statistical covariance matrix from $\chi^2$-minimization from full-spectral fitting represents the full correlated uncertainties from photon noise. In the same vein, classical fitting techniques have advantages compared to deep learning inferences like {\sc AstroNN} \citep{Leung2019a} or {\sc starnet} \citep{Fabbro2018}, as the latter does not have easy access to evaluate the observational covariance from first principles.

Alternatively, the repeated spectrum technique already used to infer {\sc aspcap}'s statistical uncertainties (\S 5.4 of \citealt{Jonsson2020}) could be extended with larger repeat samples to estimate abundance uncertainty correlations and to better capture effects that lead to non-Gaussian deviations. Our study highlights the importance of collecting repeated spectra to characterize the photon noise uncertainty. We caution that while the dispersion of abundances in star clusters is sometimes used as an empirical estimate of observational uncertainties, such an estimate is inadequate for modeling the photon noise uncertainty. The dispersion from open clusters includes both photon noise uncertainties and differential systematics, and for some elements the photon noise uncertainty could be further inflated by dispersion of intrinsic abundances within the cluster.

One of the important outcomes of large multi-element data sets has been the emergence of data-driven abundance pipelines such as {\sc the cannon} \citep{Ness2015}, {\sc AstroNN} \citep{Leung2019a}, {\sc starnet} \citep{Fabbro2018} which use labels determined from a subset of spectra to train a model that infers these labels from other spectra. This approach is especially powerful for cross-calibrating surveys with different wavelength ranges or spectral resolution \citep{Ness2015,Nandakumar2020}, and it may achieve smaller statistical uncertainties than a strictly forward modeling approach. However, a data-driven method carries the risk of ``learning'' the astrophysical correlations of abundances in stars, so that the abundance of an element is to some degree inferred from the abundances of other elements rather than from true spectral features of that element (see, e.g., fig. 19 in \citealt{Wheeler2020}, and fig. 3 in \citealt{Ting2017b}).  Under ideal conditions (e.g., noiseless labels), a generative machine learning model with the correct capacity will not be biased by correlations in the data, in contrast to a discriminative machine learning model (like {\sc AstroNN}).  However, practical data samples do not fully satisfy these ideal conditions.  Various approaches have been attempted to mitigate the effect of ``learned correlations'' with L1 regularization \citep{Casey2016}, censoring pixels, or imposing theoretical priors \citep{Ting2017b,Xiang2019}, but these mitigations are not perfect.

For the problem of measuring residual abundance correlations, the focus of this paper, a data-driven pipeline may be less well suited than a traditional forward modeling pipeline. If abundance values are partly inferred through astrophysical correlations with conditioning elements, residual correlations with these elements will be suppressed (see Fig.~\ref{fig8}), and correlations among non-conditioning elements could be artificially enhanced. We investigate this issue in Fig.~\ref{fig13}. We crossmatch our training set with the APOGEE DR14 {\sc cannon} catalog (APOGEE does not have an official DR16 {\sc cannon} catalog). This cross-matching leaves us with $13{,}207$ stars as the training set. As a comparison, we also run the same analysis using {\sc the payne}. Since the DR14 {\sc payne} catalog \citep{Ting2019b} did not measure a few elements compared to {\sc aspcap}, we re-run {\sc the payne} on DR16 using an improved set of Kurucz models \citep{Kurucz1993, Kurucz2005, Kurucz2013}, which are auto-calibrated with a machine learning technique known as domain adaptation in {\sc cycle-starnet} \citep{OBriain2021}. In order to have a more robust comparison, we train conditional normalizing flows for the {\sc aspcap}, {\sc cannon}, and {\sc payne/cycle-starnet} abundances with this common subset of $13{,}207$ stars.

Fig.~\ref{fig13} compares the results. Since {\sc the cannon} did not measure Cu, and {\sc the payne}/{\sc cycle-starnet} cannot provide robust measurement for K, we omit these two elements in this comparison. The figure shows that {\it ab initio} fitting techniques ({\sc aspcap} and {\sc the payne}) give qualitatively similar correlation structures and strong correlations ($\rho =0.4-0.5$), though there are a few notable differences. For example, {\sc the payne} abundances seem to show smaller residual correlations associated with O and stronger correlations associated with V. Some differences are not unexpected due to the different spectral models adopted\footnote{As already seen in \citet{Ting2019b}, oxygen abundances from {\sc the payne} appear to follow an ${\rm [O/Fe]}-{\rm [Fe/H]}$ trend closer to the one from the optical surveys (see their fig 12), in which ${\rm [O/Fe]}$ continues to increase for lower ${\rm [Fe/H]}$, whereas the ${\rm [O/Fe]}$ from {\sc aspcap} plateaus at low metallicity (Fig.~\ref{fig4}).}.

{\sc the cannon} abundances, by contrast, show visibly weaker residual correlations, with some qualitatively different correlation structures, despite that fact that {\sc the cannon} was trained on a higher quality subset of {\sc aspcap} abundances and therefore should have inherited the same model systematics. The differences are particularly prominent for entries that are far from the diagonal. These differences tentatively suggest that {\sc the cannon} is artificially damping some residual correlations because its abundance values are affected by ``learned'' correlations with the conditioning elements Fe and Mg. More generally, the impact of correlated measurement uncertainties on observed correlation patterns may be more difficult to evaluate for a data-driven pipeline than for an {\it ab initio} forward modeling pipeline. We plan to investigate this issue more fully in a forthcoming paper.

\begin{figure*}
    \centering
    \includegraphics[width=1.0\textwidth]{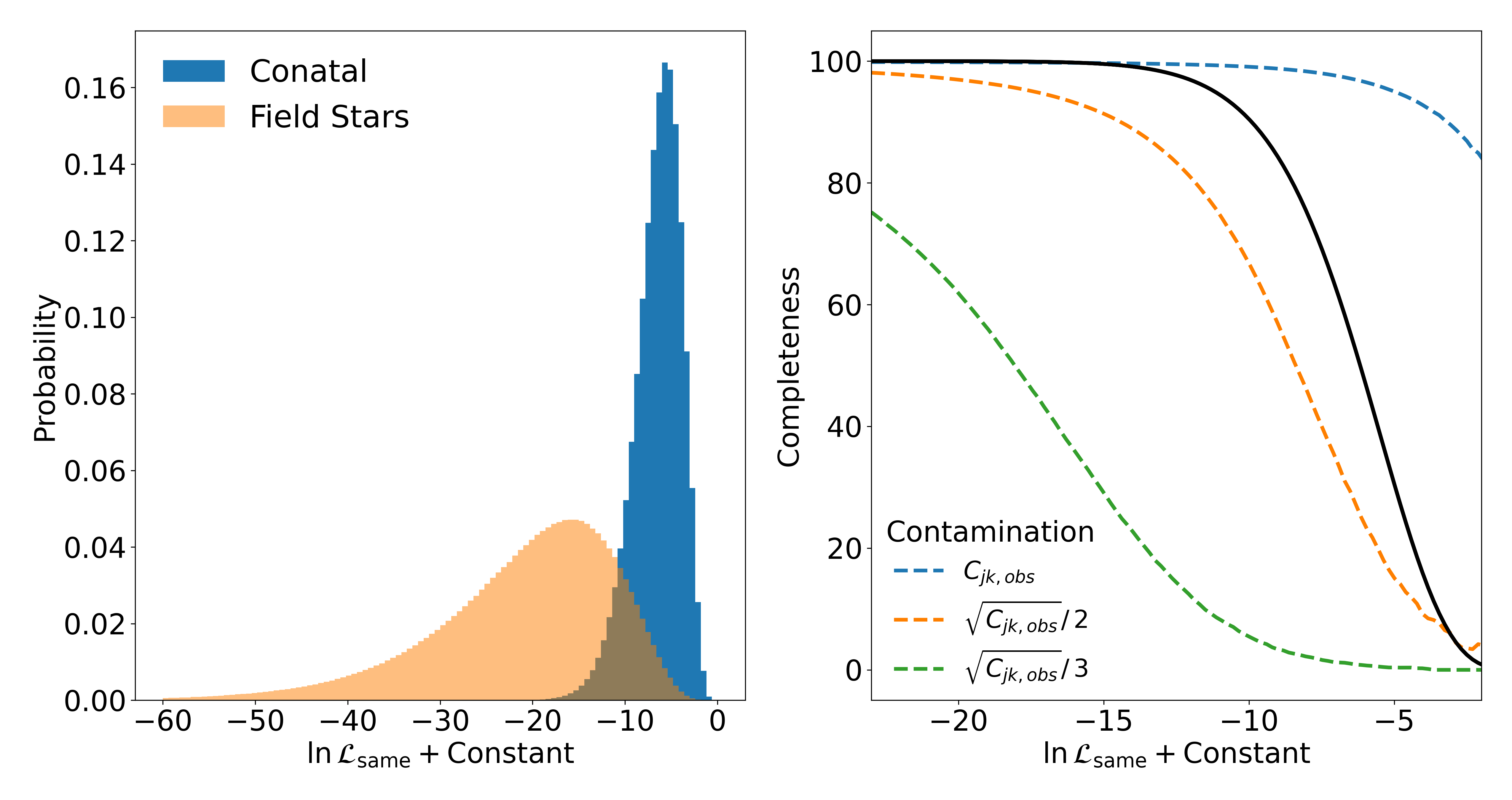}
    \caption{Reassessing chemical tagging and the chemical doppelganger rate. ({\it Left}) The blue histogram shows the probability that a pair of co-natal stars with identical intrinsic abundance has log-likelihood $\ln \mathcal{L}_{\rm same}$ (Eq.~\ref{eqn:Lsame}), assuming observational uncertainty equal to the reported {\sc aspcap} values. The orange histogram shows the log-likelihood distribution for pairs of stars drawn randomly from $\mathcal{N}(0,2\Cjktot)$, i.e., from a Gaussian approximation to the residual abundance PDF conditioned on $\feh=\mgfe=0$. ({\it Right}) The solid black curve shows the completeness of selecting co-natal pairs as a function of the adopted threshold in $\ln \mathcal{L}_{\rm same}$. The dashed blue curve shows the corresponding contamination by unrelated random pairs, assuming a field-to-co-natal ratio $f=1000$. Orange and green dashed curves show the contamination if the observational dispersion is reduced by a factor of two or three, respectively; in each case we have added a constant to $\ln \mathcal{L}_{\rm same}$ so that the completeness (black curve) stays the same. Chemical tagging with APOGEE abundances is difficult given current abundance precision and $f \simeq 1000$, but factors of 2-3 improvement in precision would make it possible to recover co-natal pairs with high fidelity.}
    \label{fig14}
\end{figure*}

%
%
%
%
%
%

\subsection{Chemical outliers, chemical siblings, and chemical doppelgangers}

Describing the elemental abundances of stars with the conditional normalizing flow and minimizing systematics through parameter conditioning offers exciting new prospects in stellar population and Galactic evolution studies. One such opportunity is a more sensitive and robust approach to finding chemical outliers, stars whose unusual abundances may reveal rare astrophysical processes or Galactic events. Very metal-poor or metal-rich stars are already rare in the local disk, but this in itself does not imply that their abundance {\it patterns} are unusual. The conditional PDF allows us to ask whether a star's abundances are unusual relative to stars of the same overall metallicity and $\alpha$-enhancement, and conditioning on $\teff$ and $\logg$ means that these unusual abundances are unlikely to be caused by differential systematic uncertainties.

Several recent studies have searched for {\it groups} of chemical outliers with similar, distinctive abundance patterns. For example, \cite{Price-Jones2020} applied DBSCAN to the 8D elemental abundance space in APOGEE and found 21 candidate (disrupted) clusters that have more than 15 members with the same ages. \cite{Ting2016} constrained the maximum mass of Milky Way star clusters through the {\it non}-detection of large, chemically homogeneous groups in APOGEE DR12. \cite{Hogg2016} applied a k-means partitioning of 15-element abundance space to APOGEE abundances derived by {\sc the cannon}, recovering multiple known star clusters through a blind search of abundance space. \cite{Ratcliffe2020} applied a hierarchical clustering algorithm to a set of 19 abundances for red clump stars in APOGEE DR14, finding that groups defined by abundances are spatially separated as a function of age. However, to date most searches have mostly found rather obvious structures, such as globular clusters and the metal-poor high-$\alpha$ population \citep[but see][]{Price-Jones2020}, or reconstructing tidal streams of known star clusters \citep[e.g.][]{Kos2018,Simpson2020}. A critical obstacle is that the sampling density of stars changes drastically with $\feh$ and $\afe$, and these changes can overwhelm more subtle signals in a clustering algorithm. Searches in the residual abundances after conditioning on Fe and Mg will be a more sensitive probe of disrupted structures with distinctive abundance patterns.

The most ambitious extension of this approach to Galactic Archaeology is the idea of ``chemical tagging,'' that we can identify pairs or groups of stars that were born in the same cluster because they have nearly identical abundances, even if they are now widely separated in phase space \citep{Freeman2002}. Since a star's atmospheric abundances do not change much in its lifetime, they can serve as the ``DNA fingerprints'' of the star's origins. Critical to the viability of this program is the ability to identify ``chemical siblings'' that have the same intrinsic abundances in the presence of ``chemical doppelgangers'' that are unrelated but have abundances that are consistent within observational uncertainties. Some recent analyses of this challenge have painted a rather bleak outlook for the ``strong chemical tagging\footnote{Unfortunately, the word ``chemical tagging'' has been used in other contexts nowadays and therefore can be confusing at times. For example, ``weak'' chemical tagging often refers to inferring the stars' birth Galactocentric radii from their elemental abundances (and ages). Here we are referring to chemical tagging in its original form -- i.e., reconstructing disrupted star clusters.}'' \citep{Ting2015,Ness2018}. Our methods and findings have numerous implications for the prospects of chemical tagging. We briefly address several of them here, reserving a detailed discussion for future work.

First, we note the obvious point that if all abundances were in fact predictable from ${\rm [Fe/H]}$ and ${\rm [Mg/Fe]}$ then the task of chemical tagging would be hopeless, as the information encoded by elemental fingerprints would be too limited. Our detection of numerous significant correlations at the measurement precision already achieved by APOGEE is an encouraging demonstration that the information content of multi-element abundances is rich. Second, our analysis shows that APOGEE really is achieving the high precision implied by repeat stellar observations, with uncertainty of 0.01-0.02 dex for eight or more elements (Fig.~\ref{fig6}). Larger observational dispersion suggested by some previous studies may be caused by differential systematic uncertainties across the sample. Third, the normalizing flow technique may be a powerful tool for chemical tagging because it can mitigate differential systematics through conditioning and because it offers a well defined way of assessing the probability of observing a given set of abundances for a star or pair of stars, including non-Gaussian features of the PDF. Finally, the correlations revealed by our APOGEE analysis alter the efficiency of distinguishing siblings from doppelgangers, an effect not previously accounted for in these estimates \citep{Ness2018}.

To calculate this last effect, we start with the two-element conditional PDF $p(\xh | \teff , \logg, \feh, \mgfe)$, evaluated at our usual reference point of Solar metallicity, $\teff = 4500\,$K, $\logg = 2.1$. Recall that the residual variances and correlations are largely independent of this choice. As discussed in \S\ref{sec:measuring_covariance}, the total covariance matrix $C_{jk,{\rm tot}}$, which we can evaluate from the conditional normalizing flow, is the sum of the observational dispersion $C_{jk,{\rm obs}}$ and the intrinsic covariance $C_{jk,{\rm int}}$. For this calculation we approximate $\Cjkobs$ as a diagonal matrix with entries given by the reported {\sc aspcap} photon noise (dashed line in Fig 6). The difference of $C_{jk,{\rm tot}}$ and $C_{jk,{\rm obs}}$ gives us the intrinsic covariance $C_{jk,{\rm int}}$.

Suppose that we observe a pair of stars whose intrinsic abundances are identical. If the observed abundance vectors $\vec{x}_1$ and $\vec{x}_2$ have Gaussian observational uncertainties, then the PDF of the abundance differences $\vec{x}_1 - \vec{x}_2$ follows a normal distribution $\mathcal{N}(0,2 C_{jk,{\rm obs}})$, with a log-likelihood
\begin{equation}
    \ln {\mathcal L}_{\rm same} = \ln \mathcal{N}(0,2 C_{jk,{\rm obs}})~.
    \label{eqn:Lsame}
\end{equation}For a pair of unrelated stars, the PDF of $\vec{x}_1 - \vec{x}_2$ instead follows $\mathcal{N}(0,2 C_{jk,{\rm tot}})$. Recall that we have already conditioned on Fe and Mg, so we are considering pairs of stars for which these two abundances are indistinguishable. The normalizing flow gives us the full and potentially non-Gaussian $p(\vec{x}_1 - \vec{x}_2)$, but here we approximate the residual abundance distribution as Gaussian to facilitate our comparisons of alternative cases below.

The left panel of Fig.~\ref{fig14} plots the distribution of $\ln \mathcal{L}_{\rm same}$ (Eq.~\ref{eqn:Lsame}) for pairs of stars that are co-natal and thus have identical abundances, and for pairs of stars drawn randomly from the conditional PDF. As expected, co-natal stars are much more likely to have small $\vec{x}_1 - \vec{x}_2$ and thus high $\ln \mathcal{L}_{\rm same}$. For chemical tagging one wants to find a threshold in $\ln \mathcal{L}_{\rm same}$ that selects most co-natal pairs while rejecting most random ``doppelganger'' pairs.

Whether or not the chemical doppelgangers will overwhelm the chemical siblings depends critically on the field-to-co-natal ratio $f$, i.e., the prior knowledge of whether a pair of stars is random as opposed to being co-natal. As discussed at length by \citet{Ting2015}, for a given star-forming event with cluster mass $M_c$, this ratio can be approximated to be $\simeq M_*/M_c$, where $M_*$ is the integrated star formation rate (SFR) of a certain Galactocentric annulus that produces stars with similar Fe and Mg to the cluster. If we further assume that the {\em total} integrated SFR at a given Galactic annulus is about $5 \times 10^{10} {\rm M}_\odot$ \citep[e.g.,][]{Ting2015,Bovy2013,Bland-Hawthorn2016}\footnote{Note that, the integrated SFR is not exactly the current stellar mass, but this ratio is roughly compensated if we only count stars within a certain Galacto-centric annulus like the Solar neighbourhood \citep[for details, see][]{Ting2015}.}, and approximate that we can grid the Fe and Mg abundance space to 1000 bins (assuming Fe and Mg photon noise precision of 0.01 dex with 0.5 dex span in ${\rm [Fe/H]}$, and 0.2 dex span in ${\rm [Mg/Fe]}$), then $f = M_c/(5\times 10^7 {\rm M}_\odot)$. For a large cluster with $M_c = 5 \times 10^4 {\rm M}_\odot$, like Westerlund 1, we will have $f \simeq 1000$, the value adopted below.

The solid black line in the right panel of Fig.~\ref{fig14} shows the completeness of selecting co-natal pairs as a function of the log-likelihood threshold, i.e., if we only consider pairs with $\ln \mathcal{L}_{\rm same}$ larger than the value on the $x$-axis. The blue dashed line shows the contamination rate, defined as the ratio of random field pairs above the threshold to the total number of pairs above the threshold, assuming a field-to-co-natal ratio of $f=1000$. With the current {\sc aspcap} photon noise illustrated in Fig.~\ref{fig6}, which corresponds to $\Delta \feh \simeq 0.01\,$dex and similar or larger values for other elements, it remains challenging to identify co-natal stars. For any threshold that has reasonable completeness, the contamination rate is nearly 100\%, i.e., doppelgangers far outnumber siblings.

However, moderate improvements in measurement precision can dramatically change this picture. Orange and green dashed curves show forecasts in which we reduce the observational dispersion by a factor of two or a factor of three (equivalently, reduce the variance by a factor of four and nine), setting $C_{jk,{\rm tot}} = C_{jk,{\rm int}} + C_{jk,{\rm obs}}/4$ and $C_{jk,{\rm int}} + C_{jk,{\rm obs}}/9$. We add a constant to the log-likelihood such that, in all three cases, the completeness results (solid black line) coincide with each other. With a factor two reduction of dispersion, one can choose a threshold that yields 80\% completeness with $\sim 50\%$ contamination, enough to yield strong statistical conclusions even if any given pair has a significant chance of being random. For a factor of three reduction of dispersion, one can choose a threshold that yields high completeness and minimal contamination. In the SNR-limited regime, these reductions would require factors of four or nine increase of observing time per star. More importantly, the {\sc aspcap} photon noise estimates, based on repeat spectra of stars, are usually larger than those estimated from $\chi^2$-fitting \citep{Jonsson2020} by a factor of a few, which implies that an abundance extraction pipeline that corrected for additional observational effects might achieve significantly higher abundance precision even with the existing APOGEE spectra. Conversely, without such corrections, simply increasing exposure times may not improve precision as rapidly as $\sigma^2 \sim 1/{\rm SNR} \sim 1/t$. Achieving sub-0.01 dex precision, even differentially, requires an even better understanding of the abundance extraction pipeline, which is no doubt challenging.

We also caution that our forecast has uncertainties because our decomposition of the observed $\Cjktot$ into $\Cjkobs$ and $\Cjkint$ can be affected by systematic uncertainties in the characterization of $\Cjkobs$. A more careful investigation of the photon noise characterization is critical (\S\ref{sec:discussion_uncertainties}). Regardless, the forecast in this study is undoubtedly a conservative estimate -- supplementing our current set of APOGEE elements with elements expected to have a larger contribution from AGB stars or other neutron-capture processes could substantially improve chemical tagging by adding more chemical variations. Chemical tagging may also be feasible in lower density regions of the residual abundance space, even if doppelgangers overwhelm siblings in the core of the distribution.  Finally, besides characterizing $\Cjkobs$, measuring a robust $\Cjktot$ as we have done in this study is also important. These correlations change the chemical tagging effectiveness because they ``tilt'' $\Cjkint$ relative to the nearly diagonal $\Cjkobs$. We repeated our $\sqrt{\Cjkobs}/3$ forecast after setting the off-diagonal elements of $\Cjkint$ to zero, and we found that ignoring the correlations can artificially alter the expected contamination rate (by 10-20\%).

\begin{figure*}
  \centering
  \includegraphics[width=1.0\textwidth]{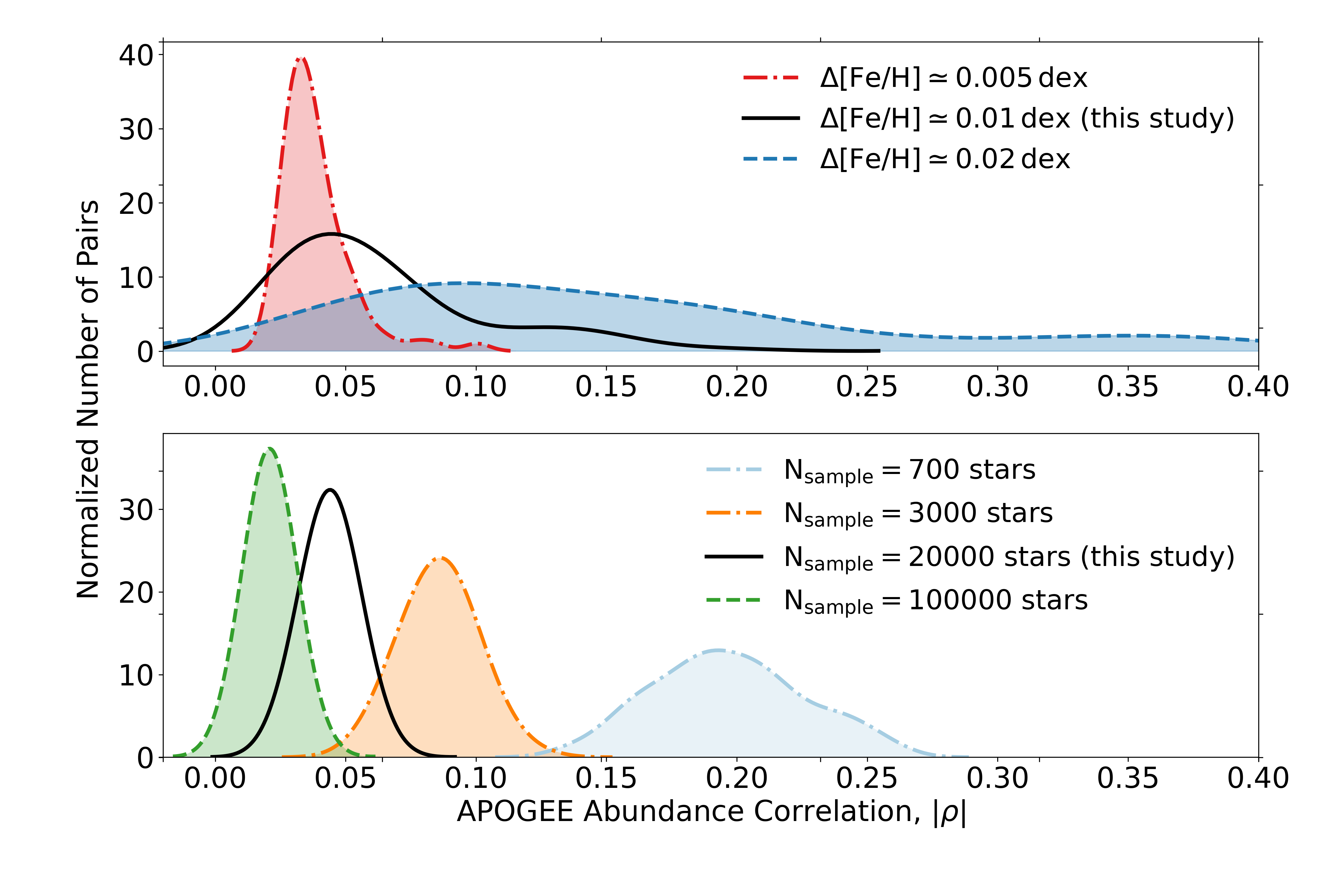}
  \caption{Influence of measurement precision and sample size on ability to detect low-amplitude correlations of residual abundance variations. ({\it Top}) Distribution of correlation coefficients produced by measurement aberration in our data set (black curve), with abundance uncertainty of $\Delta {\rm [Fe/H]} = 0.008\,$dex and $\Delta {\rm [Mg/H]} = 0.011\,$dex, and in data sets with abundance uncertainties lower (red) or higher (blue) by a factor of two. ({\it Bottom}) Distribution of the finite sampling uncertainties in correlation coefficients, as estimated from bootstrap resamplings of our APOGEE data set (black curve) and numerical experiment with samples of $700$, $3000$, or $100{,}000$ stars (blue, orange, green) from the same distribution. Although our APOGEE sample has $\simeq 20{,}000$ stars, the sampling uncertainty after conditioning on $\teff$, $\logg$, ${\rm [Fe/H]}$, and ${\rm [Mg/H]}$ corresponds to an effective sample size $N_{\rm eff} = 1/\rho^2 \simeq 500$. Detecting correlations at the level of $\rho \simeq 0.05-0.1$ requires high abundance precision and samples that are either large or targeted in stellar parameters so that $N_{\rm eff}$ approaches $N_{\rm sample}$.}
  \label{fig15}
\end{figure*}

%
%
%
%
%
%

\vspace{2cm}

\subsection{Implications for Survey Design}
\label{sec:survey-design}

The design of a Galactic spectroscopic survey involves trade-offs among numerous competing considerations, including number of targets, types of targets, spatial coverage, wavelength range, spectral resolution, and SNR. Design decisions depend partly on the instrumentation, telescope facilities, and observing time available, on synergy with other data sets, and on the prioritization of the survey's science goals. Here we discuss implications of our results for a survey that places high priority on mapping the correlations among elemental abundances, to understand nucleosynthesis and the astrophysical origin of the elements, to trace distinct stellar populations through space and time, and to probe complexities of Galactic chemical evolution. The key design considerations are to minimize systematic uncertainties in artificial correlations and to maximize the signal-to-noise ratio for detecting intrinsic correlations.

We first note that our analysis in \S\ref{sec:apogee_dispersion} suggests an intrinsic dispersion of 0.01-0.02 dex for the residual abundances of many APOGEE elements after conditioning on Fe and Mg. This is a useful reference value when thinking about the statistical precision of abundance measurements. Recall that from Eq.~\ref{eqn:rhojktot}, observed correlations are suppressed relative to intrinsic correlations by a factor that depends on the ratio of observational variance to intrinsic variance. Consequently, it will generally be difficult to measure low-amplitude intrinsic correlations for elements whose photon noise uncertainties are much larger than the intrinsic dispersion, or 0.02 dex. The fact that six out of the seven conditioning elements in this study have dispersions of $< 0.02\,$dex vividly illustrates this constraint. Nonetheless, we emphasize that measuring these correlations is still possible in a large sample if all sources of spurious correlations are sufficiently well characterized.

We have already discussed the issue of correlated statistical uncertainties in abundance measurements (\S\ref{sec:discussion_uncertainties}). This consideration favors higher spectral resolution to reduce the impact of blended features that drive stronger correlations. In principle, the correlations from statistical uncertainty can be predicted and subtracted, but this requires accurate knowledge of their magnitude. The larger they are, the more likely it is that systematic uncertainty in their level will be a limiting factor. Ideally one would like to derive the observational covariance ``theoretically'' from $\chi^2$ fitting and empirically from repeat spectra of stars, and demonstrate consistency between them. If these estimates are inconsistent, as they currently are for APOGEE \citep{Jonsson2020}, it means that the statistical uncertainties are not fully understood, and it indicates that improved data reduction and modeling might be able to extract higher precision abundances from the existing spectra. We emphasize yet again that it is valuable to minimize and fully characterize the photon noise in abundance measurements even if the absolute uncertainty is dominated by imperfections in the atmospheric and spectral synthesis models. These modeling systematics generally do not add dispersion or artificial correlations to the abundances derived for stars with similar properties. The latter can be attained by modeling the data with normalizing flows.

Higher measurement precision is desirable both to boost the expected correlation signal (assuming that the intrinsic correlation $\rhojkint$ is fixed by the underlying astrophysics) and to reduce the artificial correlations caused by measurement aberrations. In the top panel of Fig.~\ref{fig15}, the black solid curve shows the distribution of correlation coefficients produced by measurement aberration in our analysis, computed as described in \S\ref{sec:apogee_aberration}. This calculation assumes random uncertainty in ${\rm [Fe/H]}$ and ${\rm [Mg/H]}$ at the level of the mean {\sc aspcap} uncertainties for our sample, 0.008 dex and 0.011 dex, respectively. The red curve shows the predicted distribution if the random uncertainties for both elements are reduced by a factor of two, which drastically reduces the number of aberration-induced correlations above $|\rho| \simeq 0.06$. Conversely, doubling the ${\rm [Fe/H]}$ and ${\rm [Mg/H]}$ uncertainties (blue curve) leads to a much broader distribution of aberration correlations. In principle, the aberration correlations are a predictable mean signal, not an uncertainty, so one should be able to detect true correlations even at levels within this distribution. However, systematic uncertainty in the exact level of the photon noise uncertainty makes the predicted aberration signal uncertain. This systematic uncertainty in the aberration correlations is a major reason that we do not push beyond seven conditioning elements in our current analysis.

The other limiting factor in detecting weak correlations is sampling uncertainty, which scales as $N_{\rm sample}^{-1/2}$. Although not shown, we tested this theoretical scaling extensively with numerical experiments. For $N_{\rm sample}$ smaller than the current sample size, we subsampled the APOGEE sample. As for $N_{\rm sample}$ larger than the current sample size, we draw mock samples from the emulated APOGEE joint distribution (Fig.~\ref{fig5}). We found that the scaling is robust for $300 < N_{\rm sample} < 100{,}000$. We did not test beyond $N_{\rm sample} = 100{,}000$. The exact scaling also indirectly demonstrates that our normalizing flow parameterization adequately describes the distribution and only incurs negligible uncertainty. The results from some of these numerical experiments are shown in the bottom panel of Fig.~\ref{fig15}. The black curve shows the distribution of sampling uncertainty amplitudes estimated from bootstrap resamplings of our APOGEE training sample (\S\ref{sec:apogee_sampling}). The uncertainties are typically $\rho \simeq 0.045$, with some variation depending on the element correlation in question. Sampling uncertainty sets a statistical limit on the detectability of low-amplitude correlations for a given sample size. Green, orange, and blue curves show the distribution from our numerical experiments with $N_{\rm sample} = 100{,}000, 3000$, or $700$, respectively.

As previously discussed, the sampling uncertainty we find in our bootstrap analysis implies that the effective size of our sample after conditioning on $\teff$, $\logg$, ${\rm [Fe/H]}$, and ${\rm [Mg/H]}$ is only $N_{\rm eff} = 1/\rho^2 \simeq 500$, even though our full sample is $N_{\rm sample} \simeq 20{,}000$. Because our measured correlation signal is fairly consistent from one reference point to another (\ref{sec:correlation-other-teff}), we could average results from multiple reference points to reduce sampling uncertainty. We have experimented with this approach and obtained some reduction of sampling uncertainty when integrating over $\teff$ and $\logg$. However, since the training sample already spans a small range of $\teff-\logg$, we found that the reduction is minimal (from $\rho = 0.045$ to $0.041$). Expanding the sample beyond this $\teff-\logg$ range will risk distorting or diluting correlations due to the differential systematics of the abundances, though this effect could be mitigated by using median or mean trends to calibrate empirical corrections as a function of $\teff$ or $\logg$ (e.g., \citealt{Eilers2021,Ness2021,Weinberg2021}).

We found that we can approach the theoretical $\rho \simeq 0.01$ corresponding to $N_{\rm eff} \simeq 10^4$ if we further integrate over ${\rm [Fe/H]}$ and ${\rm [Mg/H]}$, but as demonstrated in Fig.~\ref{fig12}, different populations have subtle differences in the residual correlation structures, and hence we chose not to integrate the signals. The bottom line is that the same survey strategy as APOGEE as assumed in the bottom panel of Fig.~\ref{fig15} is not the most effective approach for this particular study. With a more careful selection (e.g., through Gaia's color-magnitude diagram) of ``stellar twins,'' with similar $\teff, \logg, \feh,$ and $\mgfe$, at any given reference point in this 4D-distribution, we could achieve what APOGEE enabled in this study with a sample of only $\mathcal{O}(1000)$ stars.

While normalizing flows could mitigate some of these limitations, nonetheless, collectively these considerations suggest that an effective strategy for mapping out multi-element abundance correlations might be to target moderate-sized samples ($N_{\rm sample} \simeq \mathcal{O}(10^4)$) of stars pre-selected in narrow ranges of $\teff$, $\logg$ at multiple reference points in ${\rm [Fe/H]}$, and ${\rm [Mg/Fe]}$, obtaining high SNR at high spectral resolution, analogous to solar twin studies \citep{Ramirez2009,Nissen2015,Bedell2018}. Choosing narrow ranges of stellar parameters at each reference point mitigates differential systematics as a source of observational dispersion, and it minimizes the difference between $N_{\rm sample}$ and $N_{\rm eff}$ for setting sampling uncertainty. High SNR mitigates the dilution of intrinsic correlations by observational dispersion, reduces measurement aberration, and reduces any systematic uncertainty from correlated statistical uncertainties.

Within a program such as GALAH or the SDSS-V Milky Way Mapper \citep{Kollmeier2017}, this goal could be achieved by targeting small subsets (e.g., $\mathcal{O}(1)\%$) of the full ($\ga 10^6$ star) samples for repeated observations to build high SNR. Coverage of a wide range of elements could be achieved by observing these stars in common in both optical and infrared surveys. With well-characterized correlations at various locations in ${\rm [Fe/H]}$, and ${\rm [Mg/Fe]}$, the much larger full samples could be used to search for outlier stars and chemically distinct groups and to apply these correlations to chemical tagging.

%
%
%
%
%
%

\section{Conclusions}
\label{sec:conclusions}

High-resolution, highly multiplexed spectroscopic surveys are currently measuring 10-30 elemental abundances in samples of more than $10^5$ stars. Embedded in this multi-dimensional elemental abundance space are clues to the astrophysical sources of the elements and archaeological information about the Milky Way's history. However, the tools that we currently have to decipher the irregular distribution of stars in this high-dimensional space remain rather rudimentary. In this study, we have proposed a new method based on a machine learning technique known as normalizing flow to depict the distribution of Milky Way disk stars in the abundance space spanned by 15 elements measured by the SDSS APOGEE survey. Our key findings are summarized as follows.

\begin{itemize}

    \item Conditional normalizing flow allows us to minimize the impact of differential systematic uncertainties on observational dispersion in abundance measurements. After conditioning on $\teff$, $\logg$, ${\rm [Fe/H]}$, and ${\rm [Mg/Fe]}$, the residual APOGEE abundances have a total dispersion of 0.01-0.02 dex for O, Si, Ca, Ni, $0.02-0.04\,$dex for Mn, Al, Co, S, Cr, Cu, and $0.04-0.06\,$dex for K, V, and Na. These dispersions are typically 1.5-2 times higher than the photon noise uncertainties reported by {\sc aspcap} for our ${\rm SNR}>200$ sample, and the difference is plausibly explained by intrinsic dispersion with a typical amplitude of $0.01-0.02$ dex. Differentially, the observational dispersion of APOGEE's ${\rm [Fe/H]}$ and ${\rm [Mg/Fe]}$ measurements is $\simeq 0.01$ dex or better.

   \item We have argued theoretically and demonstrated empirically that abundance correlations can be measured robustly in a large data set even when the statistical uncertainties of individual abundance measurements are comparable to the intrinsic dispersion. Studying only the dispersions about the conditional means could miss many hidden dimensions in the elemental abundance space. Abundance correlations are much more effective for measuring this subtle information.

   \item Although knowledge of a star's ${\rm [Fe/H]}$ and ${\rm [Mg/Fe]}$ is sufficient to predict most of its APOGEE abundances at the $\sim 0.02$ dex level, the residual abundances show cross-element correlations at high significance. These correlations cannot be discovered if the abundances are inferred from Fe and Mg, only if they are measured independently.

   \item Even for Solar metallicity disk stars and a set of elements expected to come mainly from core-collapse and Type Ia supernovae, we must condition on at least seven elements (e.g., Fe, Mg, O, Ni, Si, Ca, and Al) to reduce residual correlations to a level consistent with observational uncertainties. Correlation patterns for $\feh = -0.5$ stars, with $\mgfe=0.05$ or $\mgfe=0.25$, are similar to those found at Solar metallicity. Our results reconcile previous findings that ${\rm [Fe/H]}$ and ${\rm [Mg/Fe]}$ accurately predict many other abundances with other studies finding 5-10 significant dimensions to the stellar distribution in abundance space: both conclusions are true. However, since the dispersion is much less sensitive than cross-correlations of elemental abundances, the lack of reduction in dispersion does not imply that elemental abundances are redundant.
\end{itemize}

We have discussed implications of our results for survey design and analysis in \S\ref{sec:discussion}. The robust statistical modeling of the elemental space mitigates  differential systematics, improves the prospects for chemical tagging of co-natal stars, and puts the concept of chemical tagging onto a firmer statistical footing. However, chemical tagging remains challenging with current abundance precision.
For detecting and characterizing the full network of correlations, we advocate high-SNR observations of intermediate-sized samples ($\sim 10^4$ stars), concentrated in stellar parameters and at specific locations in ${\rm [Fe/H]}$ and ${\rm [Mg/Fe]}$, measuring elements that probe a range of astrophysical processes. These correlations encode critical insights about nucleosynthetic sources and subtleties of chemical evolution, and advances in theoretical modeling are needed to exploit the novel information they can provide. Once accurately measured, these correlations can be applied to larger, lower SNR samples to map temporal and spatial variations and select stars with common histories.

Normalizing flows are a powerful new technique for describing complex, high-dimensional distributions. This technique has many potential applications in Galactic Archaeology, including identification of outlier stars and distinctive clusters in abundance space. In cosmology, the advent of enormous data sets drove the development of sophisticated statistical techniques to interpret them. As Galactic Archaeology surveys grow to many elements for vast numbers of stars, a similar revolution in analysis techniques is needed to exploit the rich, multi-faceted constraints they provide on stellar astrophysics, nucleosynthesis, and the history of the Milky Way.

%
%
%
%
%
%

\acknowledgments

We thank Gregory Green for sharing his codes to visualize normalizing flows (Fig.~\ref{fig3}). We thank Jon Holtzman, Charlie Conroy and Henrik J\"onsson for illuminating discussions. YST acknowledges financial support from the Australian Research Council through DECRA Fellowship DE220101520. YST is grateful to be supported by the NASA Hubble Fellowship grant HST-HF2-51425.001 awarded by the Space Telescope Science Institute. DHW acknowledges support from NSF grant AST-1909841 and from the W.M. Keck Foundation and the Hendricks Foundation at the Institute for Advanced Study. The APOGEE/Sloan Digital Sky Survey IV is funded by the Alfred P. Sloan Foundation, the U.S. Department of Energy Office of Science, and the Participating Institutions and acknowledges support and resources from the Center for High- Performance Computing at the University of Utah. We thank Robert Kurucz for developing and maintaining the spectral synthesis programs and databases and Fiorella Castelli for allowing us to use her Linux versions of the programs.

%
%
%
%
%
%

\appendix

%
%
%
%
%
%

\section{Conditioning on different $\teff$ and $\logg$}
\label{sec:correlation-other-teff}

\begin{figure*}
  \centering
  \includegraphics[width=1.0\textwidth]{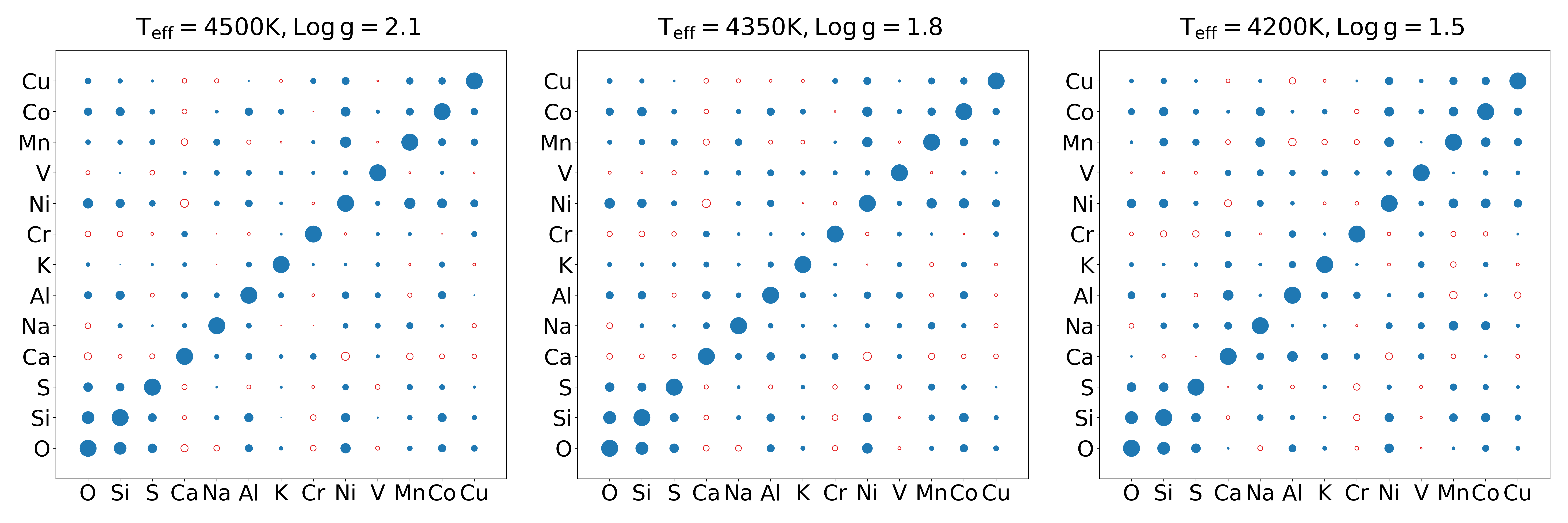}
  \caption{The abundance correlation structures at different reference $\teff$, $\logg$ when conditioning on two elements (Fe and Mg). The correlation structures, as shown in the left panel of Fig.~\ref{fig8}, are largely invariant within the stellar parameter range spanned by the APOGEE training data in this study.}
  \label{fig16}
\end{figure*}

\begin{figure*}
  \centering
  \includegraphics[width=0.9\textwidth]{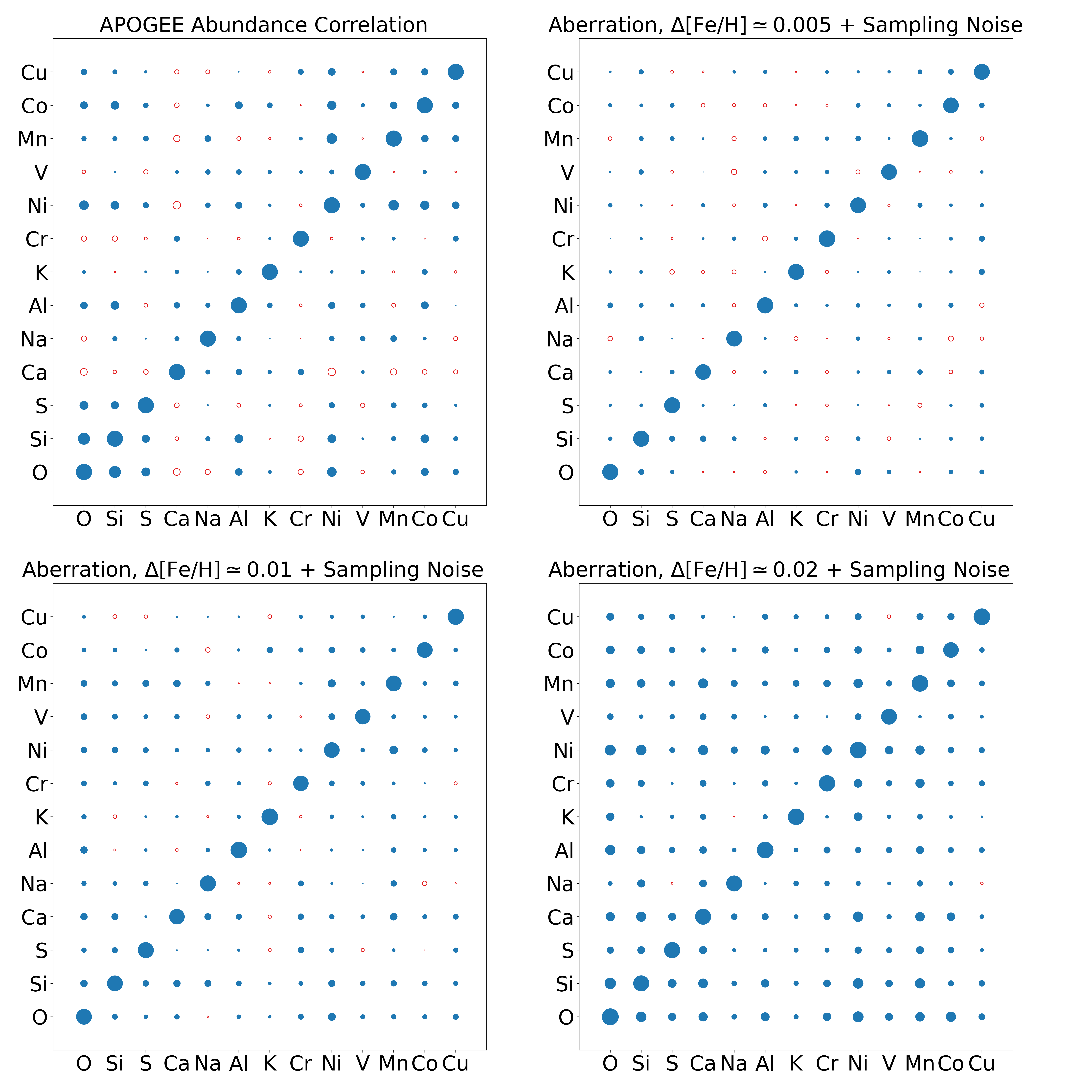}
  \caption{The measurement aberration effect, assuming different photon noise levels for {\sc aspcap}. The top left panel shows the measured APOGEE correlations (when conditioning with two elements - Fe and Mg) as a reference. The other panels demonstrate different degrees of spurious correlations due to measurement aberration adopting different photon noise uncertainties for ${\rm [Fe/H]}$ and ${\rm [Mg/H]}$. We perturb the aberration effect with the finite sampling uncertainty in the APOGEE data to better visualize the combined effect. In this study (Fig.~\ref{fig9}), we assume the {\sc aspcap} reported values -- $\Delta {\rm [Fe/H]} = 0.008$ and $\Delta {\rm [Mg/H]} = 0.011$, and the aberration effect is shown in the bottom left panel. The top right panel shows the aberration effect if the {\sc aspcap} photon noise uncertainties were two times smaller than what had been reported in DR16, and the bottom right panel two times larger. A two times larger photon noise could mimic the amplitudes of the measured signals in terms of the residual abundance correlations. However, in this case, almost all elemental abundances will be highly and positively correlated, which is at odds with the measured APOGEE abundance correlations.}
  \label{fig17}
\end{figure*}

In this study, we focus on the reference point $\teff = 4500\,$K and $\logg = 2.1$, but our results are not sensitive to this choice. Fig.~\ref{fig16} demonstrates that the residual correlation structures are almost identical across different reference $\teff$ and $\logg$ within the stellar parameters spanned by the training set.

Besides mitigating any $\teff-$abundance correlations, we found that conditioning on $\teff$ and $\logg$ is also crucial for studying the dispersions in this study. Conditioning on $\teff$ and $\logg$ reduced the measured dispersions (Fig.~\ref{fig6}). Without conditioning on $\teff=4500\,$K and $\logg=2.1$, the Mn abundances demonstrate a 70\% larger dispersion (0.037\,dex instead of 0.022 dex). The dispersions for Ca, Ni, Cr, Cu, and K also increase by 10-17\%, indicating that for some elements, abundance measurements from {\sc aspcap} have residual differential systematics even over the restricted range of $T_{\rm eff}$ in this study.

%
%
%
%
%
%

\section{Can mischaracterization of the ASPCAP photon noise explain the correlation signals?}
\label{sec:uncertainty_aberration}

In \S\ref{sec:apogee_correlations}, we have demonstrated that the {\sc aspcap} measurement correlations and finite sampling uncertainty cannot explain the APOGEE residual correlation structures. However, the measurement aberration effect, on the other hand, is sensitive to the assumption on the {\sc aspcap} photon noise. In this study, we adopt $\Delta {\rm [Fe/H]} = 0.008\,$dex and $\Delta {\rm [Mg/H]} = 0.011\,$dex as reported in {\sc aspcap}. This is likely a robust assumption because we have seen that even for elements with generally less spectral information than Fe and Mg such as O and Si, the total dispersion (the quadrature sum of the photon noise and intrinsic dispersion) is only $\sim 0.01$ dex (Fig.~\ref{fig6}).

Here we further verify this assumption by estimating the measurement aberration if {\sc aspcap} had mischaracterized the photon noise. We follow the same procedure as was done in Fig.~\ref{fig9}. The top right panel of Fig.~\ref{fig17} demonstrates the measurement aberration effect if the photon noise is two times smaller than what was reported, and the bottom right panel two times larger. The top left panel illustrates the APOGEE residual correlations as a reference, and the bottom right panel the measurement aberration in this study. Note that, unlike the bottom right panel of Fig.~\ref{fig9}, here we plot the aberration effect perturbed by the finite sampling uncertainty to visualize the combined effect of the two. Recall that the {\sc aspcap} correlation uncertainty is minimal. Therefore, the combined effect here serves as a conservative limit. Without the sampling noise, the difference between the measurement aberrations and the APOGEE correlations would be even more apparent.

The figure demonstrates that, in order to generate the strong correlations ($\rho = 0.4-0.5$) as we measured from the APOGEE data, the photon noise of {\sc aspcap} needs to be two times larger than what was reported, which is at odds with the total dispersions for the other elements. But more importantly, even in this case, as shown in the bottom right panel, the aberration effect will become so strong that all elemental abundances will be highly and positively correlated. As shown in the top left panel, the APOGEE residual correlations exhibit more structures and do not simply exhibit strong correlations among {\em all} elemental abundances. Our numerical experiment suggests that the correlations that we measured in APOGEE cannot be explained simply by the mischaracterization of the ASPCAP photon noise. This experiment also indirectly demonstrates that photon noise uncertainty reported by {\sc aspcap} is robust. Differentially, APOGEE has indeed measured ${\rm [Fe/H]}$ to a precision of 0.01 dex or better.

\bibliography{references}{}
\bibliographystyle{aasjournal}

\end{CJK*}
\end{document}